\documentstyle[12pt]{article}
\title{
{\hfill{\large IC/94/11}}\\
\ \\
FIRST QUANTIZED NONCRITICAL \\
RELATIVISTIC POLYAKOV STRING }
\author{Zbigniew Jask\'olski\thanks{
Institute
 of Theoretical Physics, Wroc{\l}aw University, pl. Maxa Borna 9,
50-204 Wroc{\l}aw, Poland; E-mail: jaskolsk@plwruw11.bitnet .}\\
Institute of Theoretical Physics, Wroc{\l}aw University, Wroc{\l}aw,
Poland\\
 \and
 Krzysztof A. Meissner\thanks{
Permanent address: Institute of Theoretical
Physics, Warsaw University, Ho\.za 69, 00-681 Warszawa, Poland;
E-mail: meissner@fuw.edu.pl . } \\
International Centre for Theoretical Physics,
 Trieste, Italy}
\date{January, 1994}
\begin{document}
\begin{titlepage}
\maketitle
\abstract{
\noindent
The first quantization of the relativistic Brink-DiVecchia-Howe-Polyakov
(BDHP) string in the range
$1<d<25$ is considered. It is shown that using the Polyakov sum over
bordered surfaces in the Feynman
path integral quantization scheme one gets a consistent quantum mechanics
of relativistic 1-dim extended objects in the range
$1<d<25$. In particular
the BDHP string propagator is exactly calculated for arbitrary
initial and final string configurations and the Hilbert space of physical
states of noncritical BDHP string is explicitly constructed.
The resulting theory is equivalent to
the Fairlie-Chodos-Thorn massive string model.
In contrast to the conventional
conformal field theory approach to noncritical string and random surfaces
in the Euclidean target space the  path integral formulation of the
Fairlie-Chodos-Thorn string obtained in this paper
does not rely on the principle of conformal invariance.
Some consequences of this feature for constructing a consistent
relativistic string theory based on the "splitting-joining"
interaction are discussed.}
\end{titlepage}
\newpage
\pagenumbering{arabic}

\section{Introduction}

In the present paper we address the question whether the Polyakov
sum over random surfaces  \cite{pol} yields in the range
$1<d<25$ a consistent
relativistic quantum mechanics of 1-dim extended objects.
Within the Feynman path integral formulation
the problem is to compute the Polyakov sum over all string
trajectories starting and ending at prescribed string configurations
and then to analyse the quantum mechanical content of the object obtained.
Our motivation for considering the problem of the first
quantized noncritical Polyakov string in this form is twofold.
First the problem has its own interest as a
nontrivial example of application
of the covariant functional quantization techniques
in the case of anomalous gauge theory.
Secondly a well developed first quantized string in the
functional formulation  may shed new light on the long-standing problem
of the interacting string theory in the physical dimension. The latter
problem has recently received a considerable attention partly stimulated
by the great progress which has been achieved over the last few years
in the noncritical Polyakov string in dimensions $d \leq 1$ and  $d=2$
\cite{gm}.
 Although there are some interesting attempts
 \cite{ag,num} to go beyond the $c=1$ barrier
 still very little is known about the Polyakov model in the
 most interesting range  $1<d<25$.

Since the existence of the consistent quantum models of the free
noncritical relativistic string is an old and well known fact
\cite{b,ct} it is probably in
order to explain what kind of new insight one
can get considering the path integral quantization of the
free Polyakov string in the position representation. For that
purpose we shall briefly consider two possible strategies of
constructing noncritical string theory based on two equivalent
but conceptually different formulations of the critical theory.
For the sake of simplicity in all further considerations
 the open bosonic string in the flat
Minkowski space is assumed.

In the so called Polyakov covariant approach the interacting
critical string theory is formulated in terms of on-shell Euclidean
amplitudes  given by the Polyakov sum over surfaces
with appropriate vertex insertions. The relativistic S-matrix in this
approach is defined by analytic continuation in the
momenta of external states. In the conformal gauge the
amplitudes can be expressed as correlators of the 2-dim
conformal field theory
(the tensor product of 26 copies  of the free scalar field)
integrated
over moduli spaces of corresponding Riemann surfaces \cite{dhph}.
In this form the Polyakov formulation can be seen as a modern covariant
version of the old dual model construction
\cite{dual} and allows for the
straightforward generalization
to arbitrary 2-dim conformal field theory with the central charge
$c=26$. In contrast to the commonly used terminology we
will call this formulation the critical Polyakov dual model.

The second independent approach starts with the quantum
mechanics of the relativistic free string.
The full interacting theory is based on the
simple picture of local "joining-splitting" interaction.
For example, the open strings can
interact by joining their end points and merging  into a
single one or by splitting a single string into two.
It is assumed that no particular interaction occurs at
joining or splitting points.
The string amplitude for a given process
is  defined as a sum over
all possible evolutions of the system. Each evolution in
this sum is represented by a world-sheet
in the Minkowski target space describing causally ordered
processes of joining and splitting and contributing
the factor ${\rm e}^{iS}$ where S is the classical string
action.
In contrast to the critical Polyakov dual model we
call this formulation the critical relativistic string
theory.

The second approach is in fact known only in the
case of the Nambu-Goto critical string in the light-cone
gauge \cite{m}.
The advantage of
this gauge is that all restrictions
imposed by the causality and locality principles of the
relativistic quantum theory can be easily implemented
in the path integral representation of the amplitudes
and the unitarity of the S-matrix is manifest.
The string amplitudes obtained within this approach
are Lorentz covariant at the critical dimension and
reproduce the  amplitudes of the old dual models.

The main difference in these approaches consists in their
fundamental organizing principles. While in the case of
the Polyakov dual model this is the principle of conformal
invariance, in the case of the relativistic string theory one starts
with the fundamental principles of the quantum mechanics of 1-dim extended
relativistic systems.

The equivalence between the critical Polyakov dual model
and the Mandelstam light cone critical relativistic
string theory, conjectured
for a long time,
has been proved few years ago \cite{gw,dg}. This is one of the
deepest and probably not fully appreciated results of
the modern string theory.
 Strictly speaking this equivalence is the only known way
 by which one can give the stringy interpretation to
 the critical Polyakov dual model and prove the
 unitarity of its relativistic S-matrix.

Let us stress that in spite of the suggestive
picture of Riemann surfaces in the Euclidean target
space (emphasized in almost every introductory text on string
theory) the relation with the world-sheets of relativistic
1-dim extended objects interacting by joining and splitting
is far from being obvious. The equivalence of the two
conceptually different
methods of constructing amplitudes in the
critical theory is based on two facts. First of all due to
the conformal invariance, the light cone diagram  can
be regarded as a special uniformization of the corresponding
punctured Riemann surface. This in particular means that the
singularities of the world-sheet in the Minkowski target space
corresponding to the joining or splitting points are inessential.
Secondly the parameters of this
diagram yield a unique cover of the corresponding moduli space \cite{gw}.
Note that the construction of the light cone diagram as well
as the range of its characteristic parameters are uniquely
determined by the causal propagation of joining and splitting strings in
the Minkowski target space. It is one of the fundamental and
nontrivial features
of the critical string theory that the basic postulates
of the relativistic quantum theory can be cast in a compact
form of modular invariance in the Euclidean formulation.

With the two equivalent formulation of the
critical string theory there are two possible ways to go beyond d=26.

The first one is to construct an appropriate generalization of
the critical Polyakov dual model
taking into account the conformal anomaly. In the covariant
continuous formulation developed in
\cite{ddk} it yields the noncritical Polyakov
dual model
given by some 2-dim conformal field theory with the central charge
$c$ coupled to the conformal Liouville theory with the central charge
$26-c$. The scheme of constructing dual amplitudes is the
same as in the critical theory - they are given by correlators
of vertex operators with the conformal weight 1. Vertices with
this property are
built from the conformal operators of the matter sector by
gravitational dressing. By construction these amplitudes are conformally
invariant.

The basic problem of the conventional  approach sketched above
is the famous $c=1$ barrier which manifests itself in the appearance
of complex critical exponents in the range $1<c<25$.
This is commonly interpreted as a manifestation of
the tachyon instability \cite{sei,ku},
although a precise mechanism of this phenomenon
in the continuum approach is still unknown.
In the case of the matter sector given by d-copies of
the free scalar fields  the noncritical Polyakov dual model
can be seen as a theory of random surfaces in the Euclidean
d-dim target space. With this interpretation it can be analysed
by random triangulation techniques
\cite{random}. The numerical simulations
suggest the branched polymer phase
\cite{num} which partly justifies
the results of the continuum approach. Also the matrix
model constructions designed to capture the range $1<c<25$
\cite{ag} indicate the polymerization of surfaces.

The second possible way to construct the interacting string
theory in the range $1<d<25$ is to follow the basic idea of
the critical relativistic string theory (in the sense assumed
in this paper).
According to the brief description given above it consists
of two steps : the relativistic quantum mechanics of 1-dim
extended objects and the derivation of scattering amplitudes
from  the simple geometrical picture of  joining-splitting
interaction. Up to our knowledge this possibility remains
completely unexplored. In fact all
the recent  achievements in noncritical string theory
rely entirely on the noncritical Polyakov dual model.

Taking the risk of missing important  results in the
rapidly developing field of research one can summarize
the current state
of affairs in the following diagram .
\vskip1cm

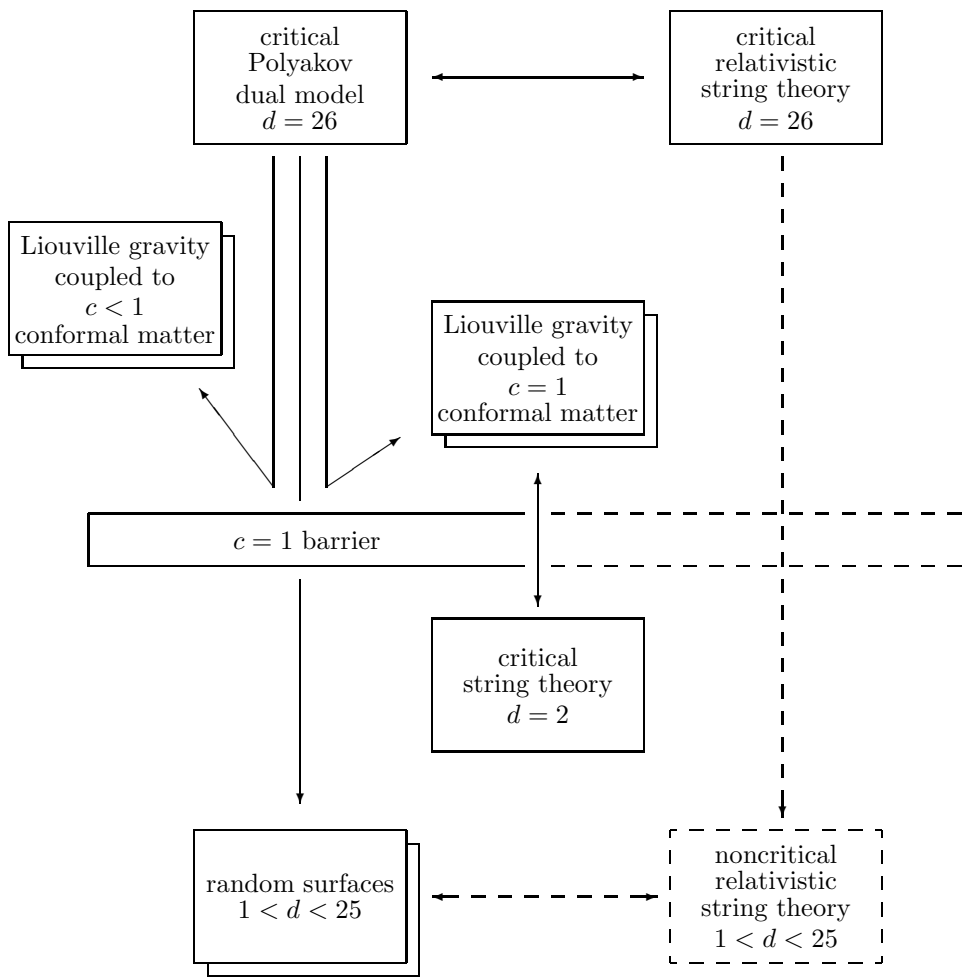
\begin{picture}(400,360)(-10,0)
\put(70,0){\framebox(80,50){\shortstack{\footnotesize
          random surfaces
          \\ \footnotesize
          $1<d<25$
          }}}
\put(75,-5){\line(1,0){80}}
\put(75,-5){\line(0,1){5}}
\put(155,45){\line(-1,0){5}}
\put(155,45){\line(0,-1){50}}
\put(250,0){\dashbox{5}(80,50){\shortstack{\footnotesize
          noncritical
          \\ \footnotesize
          relativistic
          \\ \footnotesize
          string theory
          \\ \footnotesize
          $1<d<25$
          }}}
\put(160,80){\framebox(80,50){\shortstack{\footnotesize
          critical
          \\ \footnotesize
          string theory
          \\ \footnotesize
          $d=2$
          }}}
\put(0,230){\framebox(80,50){\shortstack{\footnotesize
          Liouville gravity
          \\ \footnotesize
          coupled to
          \\ \footnotesize
          $c<1$
          \\ \footnotesize
          conformal matter
          }}}
\put(5,225){\line(1,0){80}}
\put(5,225){\line(0,1){5}}
\put(85,275){\line(-1,0){5}}
\put(85,275){\line(0,-1){50}}
\put(160,200){\framebox(80,50){\shortstack{\footnotesize
          Liouville gravity
          \\ \footnotesize
          coupled to
          \\ \footnotesize
          $c=1$
          \\ \footnotesize
          conformal matter
          }}}
\put(165,195){\line(1,0){80}}
\put(165,195){\line(0,1){5}}
\put(245,245){\line(-1,0){5}}
\put(245,245){\line(0,-1){50}}
\put(70,310){\framebox(80,50){\shortstack{\footnotesize
          critical
          \\ \footnotesize
          Polyakov
          \\ \footnotesize
          dual model
          \\ \footnotesize
          $d = 26$
          }}}
\put(250,310){\framebox(80,50){\shortstack{\footnotesize
          critical
          \\ \footnotesize
          relativistic
          \\ \footnotesize
          string theory
          \\ \footnotesize
          $d=26$
          }}}
\put(200,335){\vector(-1,0){40}}
\put(200,335){\vector(1,0){40}}
\put(110,145){\vector(0,-1){85}}
\put(100,180){\vector(-3,4){28}}
\put(120,180){\vector(3,2){28}}
\put(200,160){\vector(0,1){25}}
\put(200,160){\vector(0,-1){25}}
\put(100,180){\line(0,1){125}}
\put(110,175){\line(0,1){130}}
\put(120,180){\line(0,1){125}}

\put(30,150){\line(1,0){165}}
\put(30,170){\line(1,0){165}}
\put(30,150){\line(0,1){20}}
\put(30,150){\makebox(165,20){\footnotesize $c=1$ barrier}}

\multiput(205,150)(10,0){16}{\line(1,0){5}}
\multiput(205,170)(10,0){16}{\line(1,0){5}}

\multiput(292.5,305)(0,-10){24}{\line(0,-1){5}}
\put(292.5,65){\vector(0,-1){10}}

\put(170,25){\vector(-1,0){10}}
\put(235,25){\vector(1,0){10}}
\multiput(175,25)(10,0){6}{\line(1,0){5}}

\end{picture}

\vskip1cm
The part of the diagram drawn in solid lines corresponds to
the existing and well developed models and relations.
Shadows of some boxes indicate the matrix model versions
of the corresponding continuous models. We have also
included the so called critical 2-dim string models
arising from the interpretation of the Liouville
field in the $c=1$ noncritical Polyakov dual models
as the space-time coordinate in the target space
\cite{c1}. With this interpretation they could
be placed somewhere between the noncritical Polyakov dual model
and the noncritical relativistic string theory.

There is a number of interesting questions making the
problem of constructing the missing part of the diagram above
worth pursuing. One of them is whether the results indicating strong
instability of the noncritical Polyakov dual model (or branched
polymer nature of random surfaces) apply
to the noncritical relativistic string theory.
Clearly the answer depends on whether the equivalence between
the Polyakov dual model and the relativistic string theory
holds in noncritical dimensions. The necessary condition
for the positive answer is the conformal invariance of
the noncritical string amplitudes.
If this condition is not
satisfied or more generally if the equivalence does not
hold there is still room for a consistent relativistic
string theory in the physical dimensions, although it
is hard to expect that the amplitudes of such theory
will be dual. On the other hand if the equivalence
holds it would provide the relativistic string interpretation
of the noncritical Polyakov dual model, justifying
commonly used stringy terminology which up to now is merely
based on the equivalence in the critical dimension.

Our main motivation to the present paper was to provide the first step
toward the construction of the noncritical relativistic string theory.
The path integral
quantization of the Polyakov string in the position representation
is especially convenient for this purpose.
In fact it is the most suitable formalism for implementing the
idea of "joining-splitting" interaction. As we shall see this approach
allows for constructing the
quantum theory without the assumption of the conformal symmetry and
the equivalence mentioned above is a nontrivial and well posed problem.

The first quantized noncritical BDHP string is essentially
the problem of quantization of the anomalous theory.
Within  the Feynman path integral approach
the idea is to formulate the quantum theory entirely in
terms of the path integral over trajectories in the
configuration space without referring to the canonical
phase space analysis. In the case of model without
anomaly such formulation must of course reproduce
other methods of quantization. Applying this scheme
to the model with anomaly one may hope that the
resulting path integrals still can be given some meaning
so it will make sense to ask about the consistency
of the quantum theory derived in this way.
In particular, in the case under consideration
the conformal anomaly
manifests itself in in the appearance of the
effective action for the conformal factor. The functional measure
in the resulting
path integral can be dealt with in a similar way as in the
noncritical Polyakov dual model \cite{mm}. The difference
(and also the complication) is that
we are considering the sum over rectangular-like
surfaces connecting prescribed string configurations
in the d-dim target space which brings new effects
related with the boundary conditions.

The main result of the present paper is that
in the case of the open bosonic string described
by the BDHP action the method sketched above yields
a consistent relativistic quantum mechanics of 1-dim extended
objects. The resulting theory coincides with the 20 years old
Fairlie-Chodos-Thorn (FCT) model of the free massive string \cite{ct}.
In the radial gauge the massive string is given by the following
realization of the Virasoro algebra
\vskip.3cm
\begin{eqnarray}
L_n&=& \frac{1}{2}\sum_{m=-\infty}^{\infty} : \alpha_m\cdot
\alpha_{n-m} :
+\frac{1}{2}\sum_{m=-\infty}^{\infty} : \beta_m\beta_{n-m} :
+i(n+1) Q\beta_n
\label{bcharge}\\
\left[\beta_m,\beta_n\right]&=& m\delta_{m,-n}\;\;\;,\;\;\;
\left[\alpha^{\mu}_m,\alpha^{\nu}_n\right]\;=\; m\eta^{\mu\nu}
\delta_{m,-n}\;\;\;,\;\;\;
\eta^{\mu\nu}= diag(-1,+1,..,+1)\;\;\;.\nonumber
\end{eqnarray}
\vskip.4cm

In the Liouville sector this is the standard imaginary
background charge realization commonly used in the noncritical
Polyakov dual models \cite{pb}. The main difference consist in
the hermicity properties of the operators involved. In contrast to
the standard construction of the Feigin-Fuchs modules \cite{ff} one has
\begin{eqnarray}
\left(\alpha^{\mu}_k\right)^+ = \alpha^{\mu}_{-k}\;\;&,&\;\;
\left(\beta_k\right)^+ = \beta_{-k}\;\;\;\;\;\;k\ne 0\;\;\;,
\nonumber \\
\left(\alpha^{\mu}_0\right)^+ = \alpha^{\mu}_0\;\;&,&\;\;
\left(\beta_0 +iQ \right)^+ = \beta_0 + iQ\;\;\;\;. \label{herm}
\end{eqnarray}

As a consequence the structure of the tachyonic states is similar to that
of the critical string - there are no excited tachyonic states in the
spectrum of the free noncritical Polyakov string.

There are some points in our derivation we would like to emphasize.
First of all, in contrast to the noncritical Polyakov dual model,
the derivation is independent of the techniques
of 2-dim conformal field theory and  does not rely on
the principle of the conformal invariance.
The key point of our derivation is the exact calculation
of the transition amplitude between two arbitrary string configurations.
This is done by an appropriate extension of the space of states and
and then by translating the problem into an operator
language. The Virasoro algebra of constraints arises as a set
of consistency conditions of this method. As far as the theory of
random surfaces is concerned there is no reason for introducing the full
set of constraints. However if we assume the relativistic interpretation
of the model at hand the additional constraints acquire a physical
meaning - they can be seen as a consequence of the general
kinematical requirement which must be satisfied by wave functionals
of any quantum mechanics of relativistic 1-dim extended system.
This indicates a difference between
random surfaces and relativistic string theory. The general
(Euclidean) path integral over surfaces with fixed boundaries
is given by the model with hopelessly complicated boundary interaction.
In a sense the relativistic string theory requires a very special
type of this integral. The general theory of random surfaces
with fixed boundaries is  much more complicated and yet to be
solved problem.

The organization of the paper is as follows. In Section 2
we review the full scheme of the covariant functional
quantization of the critical string in the Schr\"{o}dinger
representation. The material included has introductory
character and serves as an illustration of the methods used
in the following. The reason for a rather lengthy form of
this section  is
twofold. First of all,
although the main idea of Feynman's quantization is well known,
we are not aware of a self-contained presentation of this technique
in the case of gauge models with reparameterization
invariance, in particular in the form suitable for
the quantization of the relativistic string. By self-contained
we mean not only path integral representation of the
transition amplitude (which is well known in the case of the critical
string propagator \cite{propa,vw})
but also the construction of the Hilbert
space of states and the derivation of the physical state conditions
from the "classical data":
the space of trajectories and the variational principle given
by the classical action.

 Secondly a large part of the analysis given in Sect.2
 applies without changes in the case of noncritical string which
 is of our main interest. This concerns in particular the
geometry of the space of trajectories (Subsect.2.1), the construction
of the space of states (Subsect.2.2) and the proper choice
of boundary conditions in the matter and in the conformal
factor sectors. The last issue is crucial for the consistent
path integral representation of the transition amplitude
(the open string propagator) constructed
in Subsect.2.3.

In Subsect.2.4 we derive the part of physical state
conditions related to the constraints linear in momenta. Within
the presented approach they are given by  generators of
the unitary realization of the residual gauge symmetry.
In Subsect.2.5 the transition amplitude between the states
satisfying constraints linear in momenta is calculated
and the second part of the physical state conditions
related to the constraints quadratic in momenta is derived.
One of these constraints - the on-mass-shell condition
is encoded in the transition amplitude. The rest can be
regarded as a consequence of the general kinematical requirement
mentioned above. Finally the
full set of the physical state conditions can be expressed
in terms of the familiar Virasoro constraints of
the old covariant formulation of the first quantized
critical string.

In Section 3 the functional formalism developed in Section 2
is applied in the case of noncritical Polyakov string
in the range $1<d<25$. The lower bound results from
the relativistic interpretation of the string which breaks
down for $d<2$ while the upper one from
the coefficient in front of the effective Liouville action
proportional to $(25-d)$. As it was mentioned above
the first few steps of the quantization procedure proceed
as in the case of the critical string. The main difference
consists in the different symmetry properties of the
noncritical string, briefly described in Subsect.3.1..

In Subsect.3.2 the transition amplitude for the noncritical
string is constructed. As a result of the conformal anomaly
and our choice of boundary conditions it involves the
path integral over conformal factor satisfying the homogeneous
Neumann boundary conditions.
As a simple consequence of the Gauss-Bonnet theorem,
the resulting
theory is stable in the case of rectangle-like world-sheets
if the bulk and the boundary cosmological
constants vanish.
The Liouville sector couples
to the "matter" sector via boundary conditions for the
$x$-variables which depend in a complicated way on the boundary
value of the conformal factor. Even with the vanishing cosmological
constant the resulting path integral cannot be directly
calculated. Our method to overcome this difficulty is to
express the transition amplitude as a matrix element of a simple
operator in a suitably extended space of states. This
is done by means of the generalized  Forman formula.

In Subsect.3.3 we derive the full set of the physical
state conditions in the extended space. It consists of
the constraints linear in momenta related to the extension itself,
the on-mass-shell condition encoded in the transition amplitude
and the set of quadratic in momenta constraints arising
by the mechanism similar to that of the critical string theory.
The resulting algebra of constraints yields the FCT
massive string model \cite{ct}. Finally in Subsect.2.4 we provide the
explicit DDF construction of the physical states of this model,
which gives a simple proof of the no-ghost theorem.

Section 4 contains the discussion of the results obtained.
A comparison with the noncritical Polyakov dual model
is given and the open problems of the
first quantized noncritical relativistic string are reviewed.
We conclude this section by a brief discussion of the choice of vanishing
cosmological constant in the interacting theory.

The paper contains three appendices. In Appendix A we gather
some basic facts concerning the corner conformal anomaly.
In Appendix B the 1-dim "conformal anomaly" is
calculated. The proof of the generalized Forman formula
is given in Appendix C.
\vspace{7mm}

\section{Functional quantization of critical Polyakov string}

\subsection{Space of trajectories}

In the Euclidean formulation a trajectory of the
open Polyakov string is given by a triplet $(M,g,x)$
where $M$ is a rectangle-like
2-dim manifold with distinguished "initial" $\partial M_i$ and "final"
$\partial M_f$ opposite boundary components, $g$ is
a Riemannian metric on $M$,
and $x$ is a map from $M$ into the Euclidean target space ${\bf R}^d$
satisfying the boundary conditions
\begin{equation}
n^a_g \partial_a x_{|\partial M_t} = 0
\label{bcxends}
\end{equation}
along the "timelike" boundary components
$\partial M_t = \partial M \setminus
(\partial M_i \cup \partial M_f)$.  In the formula above $n_g$ denotes
the normal direction along $\partial M$ with respect to the metric $g$.
In this section $d$ will be equal to 26 but will be sometimes left as $d$
to emphasize the dependence of quantities on the number of dimensions.

The BDHP action functional
$$
S[M,g,x] = \int\limits_M \sqrt{g}\, d^2z\, g^{ab}\partial_a x^{\mu}
\partial_b x_{\mu}
$$
is invariant with respect to the Weyl rescaling of the internal metric
as well as the general diffeomorphisms $f:M \rightarrow M'$ preserving
the initial and final boundary components and their orientations.
The latter invariance can be partly restricted by fixing a model manifold
$M$ and a normal direction along the boundary $\partial M$.
This can be seen as a partial gauge fixing and has important consequences
for all further constructions. Let us note that other
gauge fixings are also possible, although they are much more difficult to
deal with \cite{r}.

Let ${\cal M}_M^n$ be the space of all Riemannian
metrics on $M$ with the normal direction $n_g = n$ and ${\cal E}^n_M$
the space of all maps from $M$ into the
target space satisfying the boundary conditions (\ref{bcxends}).
In the  $(M,n)$-gauge the space of trajectories
is the Cartesian product
$ {\cal M}_M^n \times {\cal E}^n_M$.
 The gauge transformations form the semidirect product
${\cal W}_M \odot {\cal D}_M^n$ of the additive group ${\cal W}_M$
of scalar functions on $M$
and the group ${\cal D}_M^n$ of diffeomorphisms of $M$ preserving
corners and the normal direction $n$.
The action of ${\cal W}_M \odot  {\cal D}_M^n$ on
$ {\cal M}_M^n \times {\cal E}^n_M$ is given by
$$
{\cal M}_M^n \times {\cal E}^n_M
\ni (g,x)
{ \begin{picture}(90,0)(0,0) \put(5,3){\vector(1,0){80}}
\put(15,10){$\scriptstyle
(\varphi,f) \in {\cal W}_M \odot  {\cal D}_M^n
$ }
\end{picture}}
({\rm e}^{\varphi}f^*g,f^*x)
\in {\cal M}_M^n \times {\cal E}^n_M\;\;\;
$$


\subsection{Space of states}

Within the covariant functional approach to quantization
in the Schr\"{o}dinger
representation the space of states consists of wave functionals defined on
the Cartesian product ${\cal C}\times {\bf R}$
where ${\cal C}$ is a suitably chosen space
of boundary conditions (half of the Cauchy data for the classical
trajectory in the case of nondegenerate Lagrangian) and
${\bf R}$ is the time axis. For gauge
systems with the reparameterization invariance the inner time evolution is
generated by a constraint quadratic in momenta and in the subspace
of physical states it is simply given by the identity operator.
In the covariant functional
approach this feature manifests itself in  the inner time
independent formulation of variational principle.
In consequence one can describe the space of states in terms of inner time
independent wave functionals. The choice of
${\cal C}$ itself is slightly more complicated. In order to explain
the intricacies involved we consider the space
related to the reduced "position" representation in the $(M,n)$-gauge
\cite{jas}.

For every string trajectory $(g,x)\in
 {\cal M}_M^n \times {\cal E}^n_M$ we define the initial
 $(e_i,x_i)$ and the final  $(e_f,x_f)$
boundary conditions
\begin{eqnarray}
e^2_i & = &    g_{ab}t^at^b (dt)^2_{|\partial M_i}
\;\;\;,
\;\;\;(i \rightarrow f)\;\;,
\label{bc}
\\
x_i & = & x_{|\partial M_i}
\;\;\;,
\;\;\;(i \rightarrow f)\;\;\;;  \nonumber
\end{eqnarray}
where $t$ denotes a vector tangent to the boundary and $e_i, e_f$ are
einbeins induced on the initial and the final boundary component
respectively.
All possible boundary conditions form the space
$$
{\cal P}_i = {\cal M}_i  \times {\cal E}_i\;\;\;;
$$
where ${\cal M}_i$ consists of all einbeins on $\partial M_i$
and ${\cal E}_i$ is the space of maps
$x_i : \partial M_i \rightarrow {\bf R}^d $ satisfying
the Neumann boundary conditions at the ends of $\partial M_i$. Similarly
the final boundary conditions form the space
$$
{\cal P}_f = {\cal M}_f \times {\cal E}_f\;\;\;.
$$

The interpretation of the transition amplitude as an integral kernel of
some operator requires an identification of the spaces ${\cal P}_i$ and
${\cal P}_f$. It can be done by introducing a model interval $L$ and the
space
\begin{equation}
{\cal P}_L =  {\cal M}_L  \times {\cal E}_L\;\;\;
\label{sbc}
\end{equation}
together with the isomorphisms
\begin{equation}
\Gamma_i: {\cal P}_i \longrightarrow {\cal P}_L\;\;\;,\;\;\;
(i \rightarrow f) \;\;\;     ,
\label{isomor}
\end{equation}
induced by some arbitrary chosen diffeomorphisms
$$
\gamma_i: L \longrightarrow
\partial M_i \;\;\;,\;\;\; (i \rightarrow f)\;\;\;.
$$

In the covariant functional approach, the part of the canonical analysis
concerning constraints linear in momenta can be recovered by considering
classes of gauge equivalent boundary conditions. We say that $p, p' \in
{\cal P}_L$ are gauge equivalent if there exist two
${\cal W}_M \odot {\cal D}^n_M$ equivalent string trajectories with a
common
final boundary condition and starting at $p$ and $p'$ respectively.
In our case the equivalence classes can be described  as orbits of the
group ${\cal W}_L \odot {\cal D}_L$, where ${\cal W}_L$
is the additive group of real functions on $L$ satisfying,
as we shall see in the following, the Neumann boundary conditions
at the ends of $L$, and ${\cal D}_L$ is the group of orientation
preserving diffeomorphisms of $L$. The action of
${\cal W}_L \odot {\cal D}_L$ on ${\cal P}_L$
is given by
$$
{\cal P}_L \ni (e_i,x_i)
{ \begin{picture}(92,0)(0,0) \put(5,3){\vector(1,0){82}}
\put(13,10){$\scriptstyle
(\widetilde{\varphi},\gamma) \in {\cal W}_M \odot  {\cal D}_M^n
$ }
\end{picture}}
({\rm e}^{{\widetilde{\varphi}}\over 2}\gamma^*e_i,x_i \circ \gamma) \in
{\cal P}_L\;\;.
$$
The first factor in ${\cal W}_L \odot {\cal D}_L$
corresponds to the Weyl invariance in the space of
trajectories while the second one --  to
the ${\cal D}^n_M$-invariance.

In the space of states ${\cal H}({\cal P}_L)$ consisting of string wave
functionals defined on ${\cal P}_L$ the "physical" states are
${\cal W}_L \odot {\cal D}_L$ invariant functionals.
The comparison with the canonical quantization shows that the generators
of this symmetry form an operator realization of the constraints
linear in momenta. In particular the choice of the quotient
$$
{\cal K}_L = \frac{ {\cal P}_L }{ {\cal W}_L \odot {\cal D}_L }
$$
corresponds to a formulation in which all constraints linear in momenta
are solved.

As far as the constraints linear in momenta are concerned one can
construct the space of states on
an arbitrary quotient between ${\cal P}_L$ and ${\cal K}_L$.
It turns out however that the consistency requirements concerning the
path integral representation of the transition amplitude yield strong
restrictions on possible choices \cite{jas}. In general these requirements
depend on the gauge fixing used to calculate (=define) the
transition amplitude and for each particular choice of the space
${\cal C}$
boil down to some regularity conditions  concerning the gauge group
action on the space ${\cal T}[c_i,c_f]$ of trajectories starting and
ending at two fixed points $c_i,c_f \in {\cal C}$. The analysis of
the geometry of this action together with the Faddeed-Popov method
of calculating path integral can be seen as a covariant
counterpart of the symplectic reduction in the phase space approach.

In the case of the open Polyakov string in the conformal gauge there
are only two admissible choices \cite{jas}
\begin{eqnarray}
{\cal C}_L & = &  \frac{ {\cal P}_L}{{\cal D}_L }
\;\;\;,
\label{c}
\\
{\cal C'}_L & = &  \frac{ {\cal P}_L}{{\bf R}_+ \times {\cal D}_L }
\;\;\;,\nonumber
\end{eqnarray}
where ${\bf R}_+$ denotes the 1-dim group of constant rescalings
acting on ${\cal M}_L$.
Note that in the both cases above the identifications (\ref{isomor})
factor out to the canonical $\gamma$-independent isomorphisms
$$
{\cal C}_i \;= \;{\cal C}_f \;=\;{\cal C}_L \;\;\;,\;\;\;
{\cal C'}_i\; = \;{\cal C'}_f \;=\;{\cal C'}_L \;\;\;.
$$
The quotients ${\cal C}_L,{\cal C'}_L$ correspond
to the situation in which the constraints related to the Weyl invariance
are represented on the quantum level by generators of some symmetry
group acting on ${\cal C}_L$ (${\cal C'}_L$)
while all the constraints related to the ${\cal D}^n_M$-invariance
are completely solved. The only difference between the spaces
${\cal C}_L$ and ${\cal C'}_L$ consist in a different treatment of
the constraint related to the Weyl rescaling of metric by conformal factor
constant along the (initial) boundary. It is more convenient to realize
this constraint on the quantum level which corresponds to the choice of
${\cal C}_L$.

Our final task is to introduce an inner product in the space
${\cal H}({\cal C}_L)$ of string wave functionals on ${\cal C}_L$.
To this end let us observe that the space
${\cal M}_L \times {\cal E}_L$ carries the ultralocal
${\cal D}_L$-invariant structure and the scalar product in
${\cal H}({\cal C}_L)$ is given by the path integral
\begin{equation}
\langle\Psi | \Psi'\rangle =
\int\limits_{{\cal M}_L \times {\cal E}_L}
{\cal D}^ee\;{\cal D}^e\widetilde{x} \left(
{\rm Vol}_e {\cal D}_L \right)^{-1}
\overline{\Psi} \Psi'\;\;\;.
\label{product}
\end{equation}
In order to parameterize the quotient (\ref{c}) it is convenient to use
the 1-dim conformal gauge $\hat{e} = {\rm const}$ which yields
the isomorphism
$$
{\cal C}_L =_{\widehat{e}} {\bf R}_+ \times {\cal E}_L\;\;\;.
$$
Using the Faddeev-Popov method in this gauge one gets
\begin{equation}
\langle \Psi | \Psi' \rangle = \int\limits^{\infty}_0 d\alpha
\int\limits_{{\cal E}_L} {\cal D}^{\alpha \hat{e} } \widetilde{x} \;
\overline{\Psi[\alpha,\widetilde{x}]} \Psi'[\alpha,\widetilde{x}]\;\;\;.
\label{gproduct}
\end{equation}


\subsection{Transition amplitude}

The central object of the covariant functional quantization is the
path integral representation of the transition amplitude.
In the case of critical Polyakov string and with the
choice of ${\cal C}_L$ as a space of boundary conditions it takes
the following form
\begin{equation}
P[c_i,c_f] = \int\limits_{{\cal F}[c_i,c_f]} {\cal D}^gg {\cal D}^gx
\left(\mbox{Vol}_g  {\cal W}_M \right)^{-1}
\left(\mbox{Vol}_g  {\cal D}_M^n \right)^{-1}
\exp \left(- {\scriptstyle
\frac{1}{4\pi \alpha '}} S[g,x]\right)\;\;\;.
\label{pro}
\end{equation}
$ {\cal F}[c_i,c_f] \subset {\cal M}_M^n \times {\cal E}_M^n$ in
the formula above consists
of all string trajectories starting and ending at $c_i,c_f \in {\cal C}_L$
respectively, i.e. satisfying the boundary conditions
\begin{equation}
[(e_i,x_i)] = c_i \;\;\;, \;\;\; (i\rightarrow f) \;\;\;,
\label{bcx}
\end{equation}
where $e_i,x_i$ are given by (\ref{bc}) and
$[(e_i,x_i)]$ denotes the ${\cal D}_L$-orbit of the
element $(e_i,x_i) \in {\cal M}_L\times {\cal E}_L$;$ (i\rightarrow f)$.

For every $g \in {\cal M}_M^n$ the conditions (\ref{bcx})
are $g$-dependent
${\cal D}_M^n$-invariant Dirichlet boundary conditions for $x$
on $\partial M_i \cup \partial M_f$.
With the boundary conditions (\ref{bcxends}),(\ref{bcx})
the integration over $x$ in (\ref{pro}) is Gaussian
and yields ${\cal D}_M^n$-invariant functional on ${\cal M}_M^n$. This
allows for application of the F-P procedure with respect to
the group ${\cal D}_M^n$. The consistency conditions
for this method uniquely
determine boundary conditions for the metric part of a string trajectory
\cite{ja}.
The space ${\cal M}^{n*}_M$ of all metrics $g \in {\cal M}_M^n$
satisfying
these conditions can be described as follows. Let ${\cal M}^{n0}_M$
be the space of all metrics from ${\cal M}_M^n$ with the scalar curvature
$R_g = 0$, such that all boundary components are geodesic and meet
orthogonally. Then ${\cal M}^{n*}_M$ consists of all metrics
of the form $\exp (\varphi) g_0$, where $g_0 \in {\cal M}^{n0}_M$
and $\varphi$ satisfies the homogeneous Neumann
boundary condition
\begin{equation}
n^a \partial_a \varphi = 0
\label{bcphi}
\end{equation}
on all boundary components of $M$.
An important consequence of this result is that ${\cal C}_L$ is the
largest space of boundary conditions for which there exists a consistent
path integral representation of the transition amplitude.
Note that
in every conformal gauge in ${\cal M}^{n*}_M$ the conformal
factor satisfies the boundary condition (\ref{bcphi}).    It
follows that the gauge group ${\cal W}_M \odot {\cal D}^n_M$
must be restricted to the group ${\cal W}_M^n \odot {\cal D}^n_M$,
where ${\cal W}^n_M$ consists of all $\varphi \in {\cal W}_M$
satisfying the boundary conditions (\ref{bcphi}). This
justifies our definition of the induced gauge transformations
given in the previous subsection.

The boundary conditions (\ref{bcx}) yield the restrictions on the
internal length of the initial and final boundary components which
can be implemented by appropriate delta function insertions
in the path integral representation (\ref{pro}).

 With the space ${\cal F}[k_i,k_f]$
consisting of all string trajectories
$(g,x) \in {\cal M}_M^{n*} \times {\cal E}_M^d$ such that
$x$ satisfies the boundary conditions
(\ref{bcxends}),(\ref{bcx}) the calculations of the integral (\ref{pro})
proceeds
along the standard lines. In the conformal gauge
\begin{eqnarray}
(\widehat{M}_t,\widehat{g}_t)& = &([0,t]\times[0,1],
 \left( \parbox{16pt}{ \scriptsize
1 \makebox[1pt]{} 0 \\
0 \makebox[1pt]{} 1 }
 \right) ) \;\;\;,
 \label{gauge}\\
 (\widehat{L} = \partial \widehat{M}_{ti} =
 \partial \widehat{M}_{tf},\widehat{e} ) & = &
([0,t],1)\;\;\;,
\label{bgauge}
\end{eqnarray}
the $x$-integration yields
$$
\int {\cal D}^{{\rm e}^{\varphi}\hat{g}_t}x \exp (- {\textstyle
\frac{1}{4\pi \alpha'}}S[{\rm e}^{\varphi}\widehat{g}_t,x] ) =
\left( \det {\cal L}_{{\rm e}^{\varphi}\hat{g}_t} \right)^{-{d \over 2}}
\exp (- {\textstyle
\frac{1}{4\pi \alpha'}}
S[\widehat{g}_t,\varphi,\widetilde{x}_i,\widetilde{x}_f] )
$$
where
\begin{equation}
S[\widehat{g}_t,\varphi,\widetilde{x}_i,\widetilde{x}_f ] =
\int^t_0 dz^0 \int^1_0 dz^1
\left(({\partial}_0 x_{cl})^2 + ({\partial}_1 x_{cl})^2 \right)
\label{xaction}
\end{equation}
and $x_{cl}: \widehat{M}_t \rightarrow R^d$ is the solution of the
boundary value problem
\begin{eqnarray}
 \left( {\partial}^2_0 + {\partial}^2_1 \right) x_{cl} & = & 0
 \;\;,\nonumber\\
x_{cl}(0,z^1) &=& \widetilde{x}_i \circ \gamma[\widetilde{\varphi}_i](z^1)
 \;\;,
\label{xbvp}\\
x_{cl}(t,z^1) &=& \widetilde{x}_f \circ \gamma[\widetilde{\varphi}_f](z^1)
\;\;,\nonumber\\
\partial_1 x_{cl}(z^0,0) &=& \partial_1 x_{cl}(z^0,1) = 0 \;\;.\nonumber
\end{eqnarray}
The functions $\widetilde{x}_i ,\widetilde{x}_f:[0,1] \rightarrow R^d$
are representants of $c_i , c_f$ in the 1-dim conformal gauge
(\ref{bgauge})
(i.e.$ [(\alpha_i,\widetilde{x}_i)] = c_i,
[(\alpha_f,\widetilde{x}_f)] = c_f $ ),
and
the diffeomorphisms
$\gamma[\widetilde{\varphi}_i],\gamma[\widetilde{\varphi}_f] : [0,1]
 \rightarrow
[0,1]$ are uniquely determined
by the equations
\begin{eqnarray}
\frac{d}{dz^1} \gamma[\widetilde{\varphi}_i](z^1) & \propto & \exp
{\scriptstyle
\frac{1}{2}}
\widetilde{\varphi}_i(z^1) \;\;\;, \;\;\; \widetilde{\varphi}_i(z^1)
\equiv \varphi(0,z^1) \;\;,\nonumber\\
\frac{d}{dz^1} \gamma[\widetilde{\varphi}_f](z^1) & \propto & \exp
 {\scriptstyle
\frac{1}{2}}
\widetilde{\varphi}_f(z^1) \;\;\;, \;\;\; \widetilde{\varphi}_f(z^1)
\equiv\varphi(t,z^1) \;\;.
\label{difeq}
\end{eqnarray}
Solving the boundary value problem (\ref{xbvp})
and inserting solution into the action      (\ref{xaction})
one has
\begin{eqnarray*}
S[\widehat{g}_t,\varphi,\widetilde{x}_i , \widetilde{x}_f ] & = &
\frac{\left( X_{i0} - X_{f0} \right)^2}{t}
\\
& + & \frac{1}{2} \sum\limits_{m > 0}
\frac{ \pi m}{ \sinh \pi mt }
\left[ ( {X_{im}}^2 + {X_{fm}}^2 ) \cosh \pi mt -
2 X_{im}X_{fm} \right] \;\;\;, \nonumber
\end{eqnarray*}
where
\begin{eqnarray}
X^{\mu}_{i0} & = & \int\limits^1_0 dz^1
\widetilde{x}_i \circ \gamma[\widetilde{\varphi}_i](z^1)
\;\;\;\;\;\;\;\;\;\;\;\;\; (i\rightarrow f)\;\;,
\nonumber
 \\
X^{\mu}_{im} & = &2 \int\limits^1_0 dz^1
\widetilde{x}_i \circ \gamma[\widetilde{\varphi}_i](z^1) \cos \pi mz^1
\;\;\; (i\rightarrow f)\;\;.                                  \nonumber
\end{eqnarray}
Note that the functional
$S[\widehat{g}_t,\varphi,\widetilde{x}_i , \widetilde{x}_f ]$
depends only on  the boundary values $\widetilde{\varphi}_i,
\widetilde{\varphi}_f$ of
the conformal factor.

Applying the F-P method to the resulting path integral over
${\cal M}^{n*}_M$ \cite{jas}
and using the heat kernel method \cite{mm}
 to find out
the $\varphi$-dependence hidden in the functional measure and  in the
volume of  ${\cal W}_M^n$ one gets
\begin{eqnarray}
P[\alpha_f,\widetilde{x}_f;\alpha_i,\widetilde{x}_i]
= \int\limits_0^{\infty}\!\! & dt &\!\! {\eta (t)}^{1 - {d \over 2}}
t^{-{d \over 2}}
\int\limits_{{\cal W}^n}\!\! {\cal D}^{\widehat{g}_t}\varphi
\left(\mbox{Vol}_{\widehat{g}_t}  {\cal W}_M^n \right)^{-1}
\nonumber \\
&\times&\exp{\left(- {26-d \over 48\pi}
S_L[\widehat{g}_t,\varphi]\right) }
 \nonumber \\
&\times& \exp{\left( {6-d \over 32}
\sum\limits_{\mbox{\scriptsize corners}}
\varphi(z_i) \right)}
\label{propa}\\
&\times& \exp{\left(- {1 \over 4 \pi \alpha'}
S[\widehat{g}_t,\varphi,\widetilde{x}_i,\widetilde{x}_f ]
\right)} \;\;\;,\nonumber\\
&\times&
\delta \left( \alpha_f - \int_0^1 dz^1
{\rm e}^{ \frac{1}{2} \widetilde{\varphi}_f} \right) \;
\delta \left( \alpha_i - \int_0^1 dz^1
{\rm e}^{ \frac{1}{2} \widetilde{\varphi}_i} \right)
\;\;\;,\nonumber
\end{eqnarray}
where the Liouville action is given by
\begin{equation}
S_L[g,\varphi ] =
\int\limits_{M}\!\!\! \sqrt{g}\, d^2z\;
\left( {1\over 2} g^{ab} \partial_a \varphi \partial_b \varphi +
R_g\varphi
 + {\mu\over 2} {\rm e}^{ \varphi} \right)
 + \lambda \int\limits_{\partial M}\!\! e\,ds\; {\rm e}^{
\frac{\widetilde{\varphi}}{2}}
 \;\;\;,
\label{liouville}
\end{equation}
and
$$
\eta (t) = {\rm e}^{-\frac{\pi t}{12}}\prod_{n=1}^{\infty}\left(
1-{\rm e}^{-2\pi n t}\right)\;\;\;.
$$

In contrast to the expression for an on-shell open string amplitude
the conformal factor does not decouple in $d=26$.
As we shall see, the decoupling takes place for
the transition amplitudes between states in
${\cal H}({\cal C}_L)$
satisfying the constraints linear in momenta.


\subsection{Constraints linear in momenta}

The next step in the quantization procedure is to determine the
subspace of physical states in ${\cal H}({\cal C}_L)$. There are
three groups of interrelated physical state conditions:
the constraints linear in momenta given by the generators of
the residual induced gauge symmetry in ${\cal C}_L$,
the constraint quadratic in momenta encoded in the transition amplitude,
and the kinematical
constraints following from the interpretation of the string as a
1-dim extended relativistic system.
In this subsection we will discuss the first group of physical
state conditions consisting of constraints linear in momenta.

The residual gauge symmetry in the space ${\cal C}_L$ can be described by
\begin{equation}
{\cal C}_L \ni [(e,\widetilde{x})]
{ \begin{picture}(50,0)(0,0) \put(5,3){\vector(1,0){40}}
\put(15,10){$\scriptstyle
\widetilde{\varphi} \in {\cal W}_L
$ }
\end{picture}}
[({\rm e}^{{\widetilde{\varphi} \over 2}} e,\widetilde{x})]
\in {\cal C}_L\;\;\;.
\label{itrans}
\end{equation}
 In the 1-dim conformal gauge
(\ref{bgauge}) the transformation (\ref{itrans}) takes the form
\begin{equation}
{\bf R}_+ \times {\cal E}_L \ni (\alpha,\widetilde{x})
{ \begin{picture}(90,0)(0,0) \put(5,3){\vector(1,0){80}}
\put(15,10){$\scriptstyle
(\lambda,\gamma) \in {\bf R}_+ \times  {\cal D}_L
$ }
\end{picture}}
(\lambda[\widetilde{\varphi}]\alpha,\widetilde{x} \circ
\gamma[\widetilde{\varphi}])
\in \mbox{ \boldmath R}_+ \times {\cal E}_L\;\;\;,
\label{iitrans}
\end{equation}
where
$$
\lambda[\widetilde{\varphi}] = \int\limits_0^1 {\rm e}^{{\widetilde{\varphi}
\over 2}}
dz^1\;\;\;,
$$
and $\gamma[\widetilde{\varphi}]:[0,1] \rightarrow [0,1]$ is uniquely
determined
by the
equation
$$
\frac{d}{dz^1} \gamma[\widetilde{\varphi}] = \left( \lambda[\widetilde
{\varphi}]
\right)^{-1}
{\rm e}^{{\widetilde{\varphi} \over 2}}\;\;\;.
$$
It follows that all induced gauge transformations form the group
${\bf R}_+ \times {\cal D}_L$ acting on ${\cal C}_L$ by (\ref{iitrans}).
Note that this group structure as well as the group action are
consequences of the ${\cal D}^n_M$-invariant formulation and are
independent of a gauge fixing used to parameterize the quotient (\ref{c}).
This remark also applies to all further considerations where for the
sake of simplicity the conformal gauges (\ref{gauge},\ref{bgauge}) will be
used.
According to the discussion in Subsect.1.2 the wave functionals
corresponding to physical states are invariant with respect to the
induced gauge transformations represented in
${\cal H}({\bf R}_+ \times {\cal E}_L)$ by
\begin{equation}
\Psi[\alpha,\widetilde{x}]
{ \begin{picture}(90,0)(0,0) \put(5,3){\vector(1,0){80}}
\put(15,10){$\scriptstyle
(\lambda,\gamma) \in {\bf R}_+ \times  {\cal D}_L
$ }
\end{picture}}
\Psi[\lambda\alpha,\widetilde{x}\circ \gamma]
\label{naive}
\end{equation}
The problem with the representation above
is that the scalar product (\ref{gproduct}) is not invariant with
respect to (\ref{iitrans}). As a result the subspace
${\cal H}_{\scriptstyle \rm inv}({\bf R}_+ \times {\cal E}_L)$
of invariant wave functionals does not have a well defined scalar
product
and, as explained below, the transformation (\ref{naive})
should be modified.

As far as the ${\bf R}_+$ symmetry is concerned
this noninvariance is not important due to nondynamical nature of
the $\alpha$ variable. As we shall see the ambiguity in the choice
of the scalar product on the subspace
${\cal H}({\cal E}_L)$
of $\alpha$-independent
wave functionals can be hidden in the overall normalization
factor. For this reason we can restrict our considerations
to the ${\cal D}_L$ gauge symmetry in the space
${\cal H}^{\alpha\widehat{e}}({\cal E}_L)$ defined as the space
${\cal H}({\cal E}_L)$ endowed with the scalar product
\begin{equation}
\langle \Psi | \Psi' \rangle =
\int\limits_{{\cal E}_L} {\cal D}^{\alpha \hat{e} } \widetilde{x} \;
\overline{\Psi[\widetilde{x}]} \Psi'[\widetilde{x}]\;\;\;.
\label{ggproduct}
\end{equation}

 Let us recall that the functional measure ${\cal D}^e
 \widetilde{x}$ in (\ref{ggproduct}) is formally defined as the Riemannian
 volume element related to the following (weak) Riemannian structure
 on ${\cal E}_L$
$$
{ E^{e}}_{\widetilde{x}}(\delta \widetilde{x},
 \delta \widetilde{x}') =
 \int\limits_0^1 e\;ds \;\delta \widetilde{x}^{\mu}(s)
 \delta \widetilde{x}'_{\mu}(s)\;\;\;.
$$
 The pull-back of the Riemannian structure
 $E^{e}$ by the gauge transformation
 $$
 F_{\gamma} : {\cal E}_L \ni \widetilde{x}
 \longrightarrow \widetilde{x} \circ \gamma \in {\cal E}_L\;\;\;,
 $$
 is given by
$$
{F_{\gamma}^*E^{e}}_{\widetilde{x}}
(\delta \widetilde{x}, \delta \widetilde{x}') =
 \int\limits_0^1 e\;ds
 \;\delta \widetilde{x} \circ \gamma (s)
 \delta \widetilde{x}'\circ \gamma(s) =
 {E^{(\gamma^{-1})^*e}}_{\widetilde{x}}(\delta \widetilde{x},
 \delta \widetilde{x}')\;\;\;.
 $$
For $e=\alpha\widehat{e}$ where $\widehat{e}$ is a constant (as
it is in the 1-dim conformal gauge (\ref{bgauge}))
we have
 $$
(\gamma^{-1})^* e = ( \gamma^{-1})'\alpha\widehat{e}\;\;\;.
 $$
Then, calculating the 1-dim "conformal anomaly"
(Appendix B, (\ref{conano}))
and choosing the (1-dim) bulk renormalization constant equal zero one gets
$$
{\cal D}^{\alpha\widehat{e}} \widetilde{x} \circ \gamma =
{\rm e}^{-{d\over 8}(\log \gamma'(0) + \log \gamma'(1) )}
{\cal D}^{\alpha\widehat{e}} \widetilde{x} \;\;\;.
$$
It follows that in order to get the unitary ${\cal D}_L$-action
on ${\cal H}({\cal E}_L)$ the "naive" representation (\ref{naive})
of the induced gauge transformations must be replaced by
\begin{equation}
{\cal H}({\cal E}_L) \ni \Psi[\widetilde{x}]
{ \begin{picture}(50,0)(0,0) \put(5,3){\vector(1,0){40}}
\put(15,10){$\scriptstyle
\gamma \in {\cal D}_L
$ }
\end{picture}}
\rho [\gamma]
\Psi[\widetilde{x}\circ \gamma]
\in {\cal H}({\cal E}_L)\;\;\;,
\label{mtrans}
\end{equation}
where
$$
\rho[\gamma] \equiv {\rm e}^{-{d\over 16}(\log \gamma'(0) + \log \gamma'(1) )}
\;\;\;,\;\;\;\rho[\gamma \circ \delta] = \rho[\gamma]\rho[\delta]\;\;\;;
\;\;\;
\gamma,\delta \in {\cal D}_L \;\;\;.
$$
Indeed in this case we have
\begin{eqnarray*}
\int\limits_{{\cal E}_L} {\cal D}^{\alpha \hat{e} } \widetilde{x} \;
\rho[\gamma]\overline{\Psi[\widetilde{x}\circ \gamma]}
\rho[\gamma]\Psi'[\widetilde{x}\circ \gamma]&=&
\int\limits_{{\cal E}_L} {\cal D}^{\alpha \hat{e} }
\widetilde{x}\circ \gamma^{-1} \;
\rho[\gamma]^2 \overline{\Psi[\widetilde{x}]} \Psi'[\widetilde{x}]\\
&=&\int\limits_{{\cal E}_L} {\cal D}^{\alpha \hat{e} } \widetilde{x} \;
\overline{\Psi[\widetilde{x}]} \Psi'[\widetilde{x}]\;\;\;.
\end{eqnarray*}
Let us note that the modified ${\cal D}_L$-action is independent of
$\alpha$. The fully covariant with respect to the choice of $\widehat{e}$
description of the  modified action requires more general geometrical
framework and will not be discussed here.

According to (\ref{mtrans}) the space
${\cal H}_{\scriptstyle \rm inv}( {\cal E}_L)$
consists of all functionals satisfying
$$
\Psi[\widetilde{x}] =
{\rm e}^{-{d\over 16}(\log \gamma'(0) + \log \gamma'(1) )}
\Psi[\widetilde{x}\circ \gamma]\;\;\;,\;\;\;\gamma \in {\cal D}_L\;\;\;.
$$
The space ${\cal H}_{\scriptstyle \rm inv}( {\cal E}_L)$
can be also characterized in terms of constraints linear in momenta
given by the generators of the representation (\ref{mtrans})
($k\geq 1$)
\begin{eqnarray}
V_k^x \; \equiv & -& \!\!\!i\int\limits_0^1 ds \; \sin \pi ks
(\widetilde{x}^{\mu})'(s)
\frac{\delta}{\delta \widetilde{x}^{\mu}(s) }
- i \int\limits_0^1 ds \sin \pi ks  {\frac{\delta}{\delta \gamma(s)}
\rho(\gamma) }_{|\gamma = {\scriptstyle \rm id}_L }
    \label{vk}\\
 = & -& {\pi\over 2}\sum\limits_{n = 1}^k  n x_n^{\mu} p_{\mu (k-n)}
+     {\pi \over 2} \sum\limits_{n=1}^{\infty}
\left( n x_k^{\mu}p_{\mu(n+k)} - (n+k)x_{n+k}^{\mu}p_{\mu n}
\right)
  + i\pi {d\over 8}p(k)k
  \;\;\;;\nonumber
\end{eqnarray}
where
\begin{eqnarray*}
p_{\mu}(s) &\equiv & -i\frac{\delta}{\delta \widetilde{x}^{\mu}(s)}
= p_{\mu 0} + 2\sum\limits_{k = 1}^{\infty} p_{\mu k} \cos \pi ks
\;\;\;,\\
\widetilde{x}^{\mu}(s) & =
& x^{\mu}_0 + \sum\limits_{k=1}^{\infty} x_k^{\mu } \cos \pi ks\;\;\;,
\end{eqnarray*}
and
$$
p(k) = \left\{ \begin{array}{ll}
 1 & \mbox{for $k$ even} \\
0 & \mbox{for $k$ odd}
\end{array}
\right.\;\;\;.
$$
Let us note that the constraints $V^x_k$ are formally
self-adjoint operators in the
Hilbert space ${\cal H}^{\alpha\widehat{e}}({\cal E}_L)$ which
is in agreement
with the path integral derivation of the unitary ${\cal D}_L$-action
given above. Another interesting property is that $V^x_k$ are normally
ordered, a feature which is required in the canonical quantization on
different grounds.

\subsection{Constraints quadratic in momenta}

In the  covariant formulation of the first quantized relativistic particle
the Euclidean transition amplitude
is interpreted  as a  matrix element of the constraint
quadratic in momenta. Inverting this operator and performing
the Wick rotation one gets the on-mass-shell condition simply given by
the Klein-Gordon wave equation. As we shall see a similar interpretation
is valid in the case of Polyakov string.

Due to the presence of residual induced gauge symmetry it is enough
to consider the transition amplitude between states $|\Psi\rangle \in
{\cal H}({\cal C}_L)$
described by  $\alpha$-independent, ${\cal D}_L$-invariant
string wave functionals $\Psi[\widetilde{x}] $. According to
(\ref{gproduct})
and (\ref{propa}), for $d=26$ the transition amplitude between  states
$\Psi,\Psi' \in {\cal H}_{\rm \scriptstyle inv} ({\cal E}_L) \subset
{\cal H}({\cal C}_L)$ is given by
\begin{eqnarray*}
\langle \Psi | P | \Psi' \rangle &=&
\int\limits^{\infty}_0 d\alpha_f
\int\limits_{{\cal E}_L} {\cal D}^{\alpha_f \hat{e} } \widetilde{x}_f \;
\int\limits^{\infty}_0 d\alpha_i
\int\limits_{{\cal E}_L} {\cal D}^{\alpha_i \hat{e} } \widetilde{x}_i \;
\overline{\Psi[\widetilde{x}_f]} P[\alpha_f,\widetilde{x}_f;
\alpha_i,\widetilde{x}_i] \Psi'[\widetilde{x}_i]  \\
&=& \int\limits_0^{\infty}dt \; {\eta (t)}
\int\limits_{{\cal W}^n}\!\! {\cal D}^{\widehat{g}_t}\varphi
\left(\mbox{Vol}_{\widehat{g}_t}  {\cal W}_M^n \right)^{-1}
{\rm e}^{ \frac{6-d}{ 32}\!\!\!\! \sum\limits_{\mbox{
\scriptsize corners}}
\!\!\!\!\varphi(z_i) }\\
&\times&
\int\limits^{\infty}_0 d\alpha_f  \;
\delta ( \alpha_f - {\textstyle  \int
{\rm e}^{ \frac{1}{2} \widetilde{\varphi}_f} }) \;
\int\limits_{{\cal E}_L} {\cal D}^{\alpha_f \hat{e} } \widetilde{x}_f \;
\int\limits^{\infty}_0 d\alpha_i\;
\delta ( \alpha_i - {\textstyle  \int
{\rm e}^{ \frac{1}{2} \widetilde{\varphi}_i} }) \;
\int\limits_{{\cal E}_L} {\cal D}^{\alpha_i \hat{e} } \widetilde{x}_i \;\\
&\times&
\overline{\Psi[\widetilde{x}_f]}
\langle \widetilde{x}_f\circ \gamma[\widetilde{\varphi_f}] | {\rm e}^{-tH_0^x}
|
\widetilde{x}_i\circ \gamma[\widetilde{\varphi}_i]\rangle
\Psi'[\widetilde{x}_i]
\;\;\;;
\end{eqnarray*}
where
\begin{equation}
H_0^x = {\pi\over 2} \left[
 {1\over 2M_x} p^2_0 +  \sum\limits_{k=1}^{\infty} \left(
{1\over M_x} p^2_k + M_x k^2 x^2_k \right) \right] \;\;\;;\;\;\;
M_x = {1\over 4\alpha'}\;\;.
\label{ho}
\end{equation}
Changing variables
$$
\widetilde{x}_i \longrightarrow \widetilde{x}_i \circ
\gamma[\widetilde{\varphi}_i]^{-1}
\;\;\;\;,\;\;\;(i\rightarrow f)\;\;\;,
$$
and using the relations (valid for $\alpha_i =   \int
{\rm e}^{ \frac{1}{2} \widetilde{\varphi}_i}$; see Appendix B, (\ref{conano}))
\begin{equation}
{\cal D}^{\alpha_i\widehat{e}} \widetilde{x}_i^{\mu} \circ
\gamma[\widetilde{\varphi}_i]^{-1}
= {\rm e}^{ \frac{d}{16} (\widetilde{\varphi}_i(0) + \widetilde{\varphi}_i(1))}
{\cal D}^{\widehat{e}}
\widetilde{x}_i^{\mu}
\;\;\;,\;\;\;(i\rightarrow f)\;\;\;,
\label{relations}
\end{equation}
one gets for ${\cal D}_L$-invariant states
\begin{eqnarray}
\langle \Psi |P|\Psi'\rangle &=&
\int\limits_0^{\infty}dt \; {\eta (t)}
\int\limits_{{\cal W}^n_M}\!\! {\cal D}^{\widehat{g}_t}\varphi
\left(\mbox{Vol}_{\widehat{g}_t}  {\cal W}_M^n \right)^{-1}
{\rm e}^{ \frac{ 3}{  16}\!\!\!\! \sum\limits_{{ \scriptstyle {\rm corners}}}
\!\!\!\!\varphi(z_i) }
\left({\textstyle \int {\rm e}^{{1\over 2}\varphi_i} \int
{\rm e}^{{1\over 2}\varphi_f} }
\right)^{{13 \over 4}}
\nonumber\\
&\times&
\int\limits_{{\cal E}_L} {\cal D}^{ \hat{e} } \widetilde{x}_f \;
\int\limits_{{\cal E}_L} {\cal D}^{ \hat{e} } \widetilde{x}_i \;
\overline{\Psi[\widetilde{x}_f]}
\langle \widetilde{x}_f| {\rm e}^{-tH_0^x} |
\widetilde{x}_i\rangle
\Psi'[\widetilde{x}_i]\;\;\;.\nonumber
\end{eqnarray}
In the formula above the integration over conformal factor decouples
yielding an overall divergent factor independent of the states
$\Psi, \Psi' \in {\cal H}_{\rm \scriptstyle inv} ({\cal C}_L)$.
It follows that one can restrict oneself to the space
${\cal H}({\cal E}_L) \subset {\cal H}({\cal C}_L)$ of
$\alpha$-independent
states endowed with the scalar product (\ref{ggproduct}) with $e=
\widehat{e}$.
Then the (regularized) transition amplitude between
${\cal D}_L$-invariant
states in ${\cal H}^{\widehat{e}}({\cal E}_L)$
takes the following simple form
\begin{equation}
\langle\Psi|P_R|\Psi'\rangle =
\langle\Psi|\int\limits_0^{\infty} dt\; \eta(t) {\rm e}^{-tH_0^x}|\Psi'\rangle
\;\;\;.
\label{propagat}
\end{equation}

Let us note that the path integral representation (\ref{propa}) of
the transition amplitude is well defined only on the subspace
${\cal H}_{\rm \scriptstyle inv} ({\cal E}_L) \subset
{\cal H}^{\widehat{e}}({\cal E}_L)$.
In order to get a well defined representation
in the whole Hilbert space
${\cal H}^{\widehat{e}}({\cal E}_L)$ one has to restrict the
space of trajectories in (\ref{pro}) such that the conformal factor
already decouples in the formula (\ref{propa}). This restriction
depends on
some additional geometrical data which in the conformal gauges
(\ref{gauge},\ref{bgauge}) consist of fixed parameterizations of
the initial and final boundary components. Any particular choice of
this data leads to some extension of the formula (\ref{propagat})
to the space ${\cal H}^{\widehat{e}}({\cal E}_L)$ and can be regarded as a
choice
of gauge in the second quantized theory \cite{jas,jask}.
For the sake of simplicity
we will use the simplest extension
\begin{equation}
P_R[\widetilde{x}_f,\widetilde{x}_i] =
\langle \widetilde{x}_f|P_R|\widetilde{x}_i\rangle
 = \langle \widetilde{x}_f|
 \int\limits_0^{\infty} dt\; \eta(t) {\rm e}^{-tH_0^x}
 |\widetilde{x}_i\rangle \;\;\;.
\label{pr}
\end{equation}
It should be stressed that all further considerations are
independent of this choice.

In order to derive the on-mass-shell condition one would like to
invert the operator $P_R$.
The problem is that, due to the $\eta$-function
insertion, $P_R$ is not invertible on the whole space
${\cal H}^{\widehat{e}}({\cal E}_L)$.
Moreover the formula (\ref{pr}) does not describe any operator
on the subspace ${\cal H}_{\rm \scriptstyle inv} ({\cal E}_L)$
of ${\cal D}_L$-invariant states. The
largest subspace of ${\cal H}_{\rm \scriptstyle inv} ({\cal E}_L)$
on which $P_R[\widetilde{x}_f,\widetilde{x}_i]$
can be regarded as an integral kernel of a well
defined operator is characterized by the equations
\begin{equation}
H_k^x |\Psi\rangle = 0\;\;\;,\;\;\;V_k^x |\Psi\rangle = 0\;\;\;
;k=1,...\;\;\;,
\label{phsc}
\end{equation}
where
\begin{eqnarray}
H_k^x& \equiv& -{i\over \pi k} [H_0^x,V_k^x]    \label{hk} \\
&=& { \pi\over 2} \left[ {1\over 2M_x}p_k \cdot p_0 +
{1\over 2}\sum\limits_{n=1}^{k} \left( {1\over M_x}
p_n \cdot  p_{k-n}
-M_x n(k-n)
x_n \cdot x_{k-n} \right) \right.       \nonumber           \\
&+& \left. \sum\limits_{n=1}^{\infty} \left( {1\over M_x}
p_n \cdot p_{k+n} +
M_x n(k+n)x_n \cdot x_{k+n} \right) \right]\;\;\;.    \nonumber
\end{eqnarray}

In order to analyse the integrability conditions of the equations
(\ref{phsc}) it is convenient to introduce the operators
$$
L^x_{\pm k} \equiv {1\over \pi} (H_k^x \pm iV_k^x)\;\;\;, \;\;\;k=1,...\;;
$$
which are just the standard representations of
the Virasoro generators :
\begin{eqnarray}
L^x_k &=& {1\over 2} \sum\limits_{-\infty}^{+\infty} \alpha_{-n}
\cdot \alpha_{k+n} \;\;\;,\;\;\;k=\pm 1,\pm 2,...\label{lxk}\\
\alpha_0^{\mu} &\equiv& {1\over \sqrt{2M_x}} p_{\mu 0}\;\;\;,\nonumber\\
\alpha_n^{\mu } &\equiv & {1\over \sqrt{2}}
\left( {1\over \sqrt{M_x}}p_{\mu n}-i \sqrt{M_x} nx_n^{\mu} \right)
\;\;\;,\;\;\;n=\pm 1,\pm 2,...\;\;\;;\nonumber
\end{eqnarray}
$$
\left[\alpha_n^{\mu},\alpha_m^{\nu}\right] =
 n \delta^{\mu \nu} \delta_{n,-m}\;\;\;\;\;
,\;\;\;\;\;\alpha_n^{\mu } = \alpha_{-n}^{\mu}\;\;\;.
$$
In terms of $L^x_k$ the integrability conditions take the form
$$
\left[ L^x_n ,L^x_m \right] =
(n-m) L^x_{n+m} +  {26 \over 12}
\delta_{n,-m}(n^3-n)\;\;\;,
$$
where
\begin{equation}
L^x_0 \equiv {1\over 2}\alpha_0 \cdot \alpha_0 +
\sum\limits_{n=1}^{\infty} \alpha_{-n} \cdot \alpha_n\;\;\;.
\label{lo}
\end{equation}

In order to obtain nontrivial solutions to the equations (\ref{phsc})
one has to relax their  strong form. Since $L_k^{x+} = L^x_{-k}$,
this leads to the familiar conditions
for the off-mass-shell physical states
\begin{equation}
L^x_k|\Psi\rangle = 0\;\;\;,\;\;\; k\geq1\;\;\;.
\label{oph}
\end{equation}

The derivation of the physical state conditions presented above
requires an explanation.
The additional constraints $H_k^x$ have been
introduced for technical reasons. There is however a physical motivation
for these constraints stemming from the fact that the string is an
extended relativistic system. This
implies that the intersection of the string world sheet by an
equal time hyperplane provides a half of the Cauchy data for
string trajectory. It follows that for each particular choice
of reference system in the Minkowski space-time the string wave
functional should be independent of string fluctuations in the
time direction. This kinematical requirement can be regarded as
a manifestation of the general locality and causality principles
of relativistic quantum theory and indicates a fundamental
difference between the theory of relativistic string and the
theory of random surfaces. Note that in the canonical quantization
the kinematical requirement simply means that
all negative norm states must decouple.

The way in which the $H_k^x$ constraints
yield exactly the missing part of the physical state conditions
necessary and sufficient to satisfy the kinematical requirement
is not quite straightforward.
The rough counting of the degrees of freedom shows that two sets
of constraints $\left\{ V_k\right\}$ and $\left\{ H_k \right\}$
reduce by 2 the number of physical direction. However, as it was
mentioned above one can impose only a half of these constraints
as conditions for physical states, which explicitly removes
only one direction. The mechanism which ensures the decoupling
of the second direction is that of the null states which are among
solutions to the equations (\ref{oph}) \cite{re,sch}.
As the null states decouple
from all other solutions  the space of physical
states is effectively given by the space of equivalence classes.
One possible way to describe the corresponding  quotient  is
to introduce the Euclidean "quantum"
version of the light-cone gauge conditions
\begin{equation}
\alpha^+_k|\Psi\rangle \equiv \left( i\alpha_k^0 + \alpha_k^{25} \right) |
\Psi\rangle = 0\;\;\;,\;\;\;k\geq 1\;\;\;.
\label{lc}
\end{equation}
The conditions above can be seen as an explicit implementation of the
kinematical requirement in the limit case of the light-like
direction. It is an interesting open question whether the
Virasoro algebra of constraints is the only solution to the
kinematical requirement of the quantum theory of 1-dim extended
relativistic systems.

The subspace
${\cal H}_{\scriptstyle \rm ph}^{\scriptstyle \rm off}({\cal E}_L)$
of solutions to the equations (\ref{oph},\ref{lc}) can be explicitly
constructed by means of the (extended) DDF method \cite{ddf}.
One can check that the operators $\left\{ \widetilde{x}^{\perp}(s)
\right\} \equiv \left\{ \widetilde{x}^k(s) \right\}_{k=1,...,d-2}$
form in
${\cal H}_{\scriptstyle \rm ph}^{\scriptstyle \rm off}({\cal E}_L)$
a complete set of commuting "observables". In particular
every state
$|\Psi_{\scriptstyle \rm ph}\rangle \in
{\cal H}_{\scriptstyle \rm ph}^{\scriptstyle \rm off}({\cal E}_L)$
is completely described by the wave functional
\begin{equation}
\Psi_{\scriptstyle \rm ph}[\widetilde{x}^{\perp}] =
\langle \widetilde{x}^{\perp} |\Psi_{\scriptstyle \rm ph} \rangle
\;\;\;.
\label{st}
\end{equation}

The transition amplitude between off-mass-shell physical states
$\Psi_{\scriptstyle \rm ph},\Psi_{\scriptstyle \rm ph}'
 \in {\cal H}_{\scriptstyle \rm ph}^{\scriptstyle \rm off}({\cal E}_L)$
described by the wave functionals (\ref{st})
takes the following form
\begin{eqnarray}
\langle\Psi_{\scriptstyle \rm ph}|P_R|\Psi_{\scriptstyle \rm ph}'\rangle
&=&
\int\limits_{{\cal E}_L} {\cal D}^{ \hat{e} } \widetilde{x}_f \;
\int\limits_{{\cal E}_L} {\cal D}^{ \hat{e} } \widetilde{x}_i \;
\overline{\Psi_{\scriptstyle \rm ph}[\widetilde{x}^{\perp}_f]}
\langle \widetilde{x}_f|
 \int\limits_0^{\infty}\!\! dt\; \eta(t) {\rm e}^{-tH_0^x}|
\widetilde{x}_i\rangle
\Psi_{\scriptstyle \rm ph}'[\widetilde{x}^{\perp}_i] \nonumber \\
&=& \langle\Psi_{\scriptstyle \rm ph}|
 \int\limits_0^{\infty} \!\!dt\;  {\rm e}^{-t\pi (K^x_0 -1)}
|\Psi_{\scriptstyle \rm ph}'
\rangle_{\scriptstyle \rm ph}^{\scriptstyle \rm off}
\;\;\;,
\label{matrix}
\end{eqnarray}
where $K^x_0$ denotes the restriction of the operator $L^x_0$ to the
subspace
${\cal H}_{\scriptstyle \rm ph}^{\scriptstyle \rm off}({\cal E}_L)$
and $\langle...|...\rangle_{\scriptstyle \rm ph}^{\scriptstyle \rm off}$
is the scalar product in
${\cal H}_{\scriptstyle \rm ph}^{\scriptstyle \rm off}({\cal E}_L)$
regarded as a Hilbert subspace of ${\cal H}^{\widehat{e}}({\cal E}_L)$.
It follows that the operator defined by the matrix elements
(\ref{matrix}) can be inverted  on the subspace
${\cal H}_{\scriptstyle \rm ph}^{\scriptstyle \rm off}({\cal E}_L)$.
Then the on-mass-shell condition in
${\cal H}_{\scriptstyle \rm ph}^{\scriptstyle \rm off}({\cal E}_L)$
is given by
$$
(K^x_0 - 1) |\Psi\rangle = 0\;\;\;,
$$
which is equivalent to the condition
\begin{equation}
(L^x_0 - 1) |\Psi\rangle = 0\;\;\;.
\label{mshell}
\end{equation}
in the space ${\cal H}^{\widehat{e}}({\cal E}_L)$.

Performing the Wick rotation in the physical state conditions
(\ref{oph},\ref{mshell}) one gets  the familiar equations
of  the so called old covariant approach.
All further steps of quantization proceed along the
standard lines \cite{re,sch}.
\vspace{7mm}


\section{The noncritical Polyakov string}

\subsection{Gauge symmetry }

In this section we will present the covariant functional quantization
of the  Polyakov string
in the flat target space of dimension $d$ in the range $1<d<25$.
 The first steps of this quantization
procedure -- the description of the space of trajectories in the
configuration space, the construction of the space of states, and
the path integral representation of the transition amplitude --
are almost the same as in the case of the critical string. The main
difference
consists in the symmetry requirements imposed on the quantum theory.

Since, in the range $1<d<25$, the conformal anomaly breaks the Weyl
invariance
completely, the gauge symmetry in the space of trajectories reduces to the
group ${\cal D}_M^n$ acting on ${\cal M}_M^n \times {\cal E}_M^n$ by
$$
{\cal M}_M^n \times {\cal E}^n_M
\ni (g,x)
{ \begin{picture}(60,0)(0,0) \put(5,3){\vector(1,0){50}}
\put(15,10){$\scriptstyle
f \in {\cal D}_M^n
$ }
\end{picture}}
(f^*g,f^*x)
\in {\cal M}_M^n \times {\cal E}^n_M\;\;\;.
$$
Accordingly, the induced gauge transformations in the space ${\cal P}_L$
(\ref{sbc}) of boundary conditions take the form
$$
{\cal P}_L \ni (e_i,x_i)
{ \begin{picture}(50,0)(0,0) \put(5,3){\vector(1,0){40}}
\put(13,10){$\scriptstyle
\gamma \in  {\cal D}_L
$ }
\end{picture}}
(\gamma^*e_i,x_i \circ \gamma) \in
{\cal P}_L\;\;.
$$
Repeating the reasoning of Subsect.2.2 one gets the space of states
${\cal H}({\cal C}_L)$ endowed with the scalar product (\ref{product}).

Because of the restricted gauge group there is no residual gauge
invariance in ${\cal H}({\cal C}_L)$.

\subsection{Transition amplitude}

According to the different symmetry requirements,
the path integral representation (\ref{pro}) for the transition
amplitude gets slightly modified
$$
P[c_f,c_i] = \int\limits_{{\cal F}[c_f,c_i]} {\cal D}^gg {\cal D}^gx
\left(\mbox{Vol}_g  {\cal D}_M^n \right)^{-1}
\exp \left(- {\scriptstyle
\frac{1}{4\pi \alpha '}} S[g,x]\right)\;\;\;.
$$
In the conformal gauges (\ref{gauge},\ref{bgauge}) one has
\begin{eqnarray}
P[\alpha_f,\widetilde{x}_f;\alpha_i,\widetilde{x}_i]
= \int\limits_0^{\infty}\!\! & dt &\!\! {\eta (t)}^{1 - {d \over 2}}
t^{-{d \over 2}}
\int\limits_{{\cal W}^n_M}\!\! {\cal D}^{\widehat{g}_t}\varphi
\nonumber \\
&\times&\exp{\left(- {25-d \over 48\pi}
S_L[\widehat{g}_t,\varphi]\right) }
 \nonumber \\
&\times& \exp{\left( {7-d \over 32}
\sum\limits_{\mbox{\scriptsize corners}}
\varphi(z_i) \right)}
\label{npropa}\\
&\times& \exp{\left(- {1 \over 4 \pi \alpha'}
S[\widehat{g}_t,\varphi,\widetilde{x}_i,\widetilde{x}_f ]
\right)} \;\;\;,\nonumber\\
&\times&
\delta \left( \alpha_f - \int_0^1 dz^1
{\rm e}^{ \frac{1}{2} \varphi_f} \right) \;
\delta \left( \alpha_i - \int_0^1 dz^1
{\rm e}^{ \frac{1}{2} \varphi_i} \right)
\;\;\;.\nonumber
\end{eqnarray}
Lets us note that in contrast to the formula (\ref{propa}) the
$\varphi$-dependence of the functional measure
${\cal D}^{{\rm e}^{\varphi}\widehat{g}_t}\varphi$
is not canceled by the similar $\varphi$-dependence in the volume factor
$\mbox{Vol}_{{\rm e}^{\varphi}\widehat{g}_t}  {\cal W}_M^n$.
As a result, one gets  the
different coefficients in front of the Liouville
action and the  corner anomaly term.

Some remarks concerning the formula (\ref{npropa}) are in order.
First of all one has to choose some values of the renormalization
constants $\mu,\lambda$ appearing in the Liouville action
(\ref{liouville}).
In the following we will restrict ourselves to the simplest choice
\begin{equation}
\mu = \lambda = 0\;\;\;.
\label{cosm}
\end{equation}
This is well justified in the free theory. First, let us observe that
the nonvanishing boundary
cosmological constant is incompatible with the boundary conditions
(\ref{bcphi}). Secondly, using the Gauss-Bonnet theorem on the
rectangle one can easily show that for $\widehat{g}_t \in {\cal M}^{n*}_M$
the Liouville equation
$$
\Delta_{\widehat{g}_t} \varphi + \mu {\rm e}^{\varphi} =0 \;\;\;,
$$
does not have any solution in the space ${\cal W}_M^n$.
It follows that the variational problem given by the Liouville action
(\ref{liouville}) is well posed and has a solution in ${\cal W}_M^n$ if
and only if the equations (\ref{cosm}) are satisfied.
Let us stress
that the conclusion above is not necessarily
valid for more complicated world sheet
topologies. In particular in the case of hyperbolic hexagon
the classical solution exists only for $\mu > 0$. An independent
motivation
for the choice (\ref{cosm}) in the free string theory stems
from the semiclassical calculations of the static potential
\cite{jm}, where
one obtains the same result for the vanishing and for the positive
bulk cosmological constant.

Under the assumption (\ref{cosm}) and in the conformal gauge (\ref{gauge})
the Liouville action is just the free field action
$$
S[\widehat{g}_t,\varphi ] =    {1 \over 2}
\int^t_0 dz^0 \int^1_0 dz^1
\left(({\partial}_0 \varphi)^2 + ({\partial}_1 \varphi)^2 \right)
\;\;.
$$
Even with this simplification the formula (\ref{npropa})
still contains the complicated nonlocal interaction
which prevents calculations of the functional integral in all but few
special cases of $x$-boundary conditions \cite{jm}. Our idea to overcome
this difficulty is to regard  the transition amplitude (\ref{npropa}) as
a matrix element of some simple operator  between
special states in an  extended space.
The  method to find such extension of ${\cal H}({\cal C}_L)$ is based on
the simple
observation that the interaction terms depend only on the boundary values
of
the conformal factor. Thus it should be possible to replace the
integration
over fields satisfying the homogeneous Neumann boundary condition
by the integral over fields with a fixed nonhomogeneous Dirichlet
condition
(which is Gaussian) and then an integral over all possible boundary
values.
In the case under consideration one can expect  the following formula
\begin{equation}
\int\limits_{{\cal W}^n_M} \!\!\! {\cal D}^{\widehat{g}_t}\varphi\;
{\rm e}^{ - {25-d \over 48\pi} S_L[\widehat{g}_t,\varphi] }
F[\widetilde{\varphi}_f,\widetilde{\varphi}_i]
= \eta^{-{1\over2}} t^{-{1\over 2}}
    \int\limits_{{\cal W}_L} \!\!\!
    {\cal D}^{\widehat{e}} \widetilde{\varphi}_f
    \int\limits_{{\cal W}_L} \!\!\!
    {\cal D}^{\widehat{e}} \widetilde{\varphi}_i
\;{\rm e}^{ - {25-d \over 48\pi}
S_L[\widehat{g}_t,\widetilde{\varphi}_f,\widetilde{\varphi}_i] }
F[\widetilde{\varphi}_f,\widetilde{\varphi}_i]\;\;,
\label{formula}
\end{equation}
where
$$
S_L[\widehat{g}_t,\widetilde{\varphi}_f,\widetilde{\varphi}_i] \equiv
S_L[\widehat{g}_t,\varphi_{cl}]\;\;\;,
$$
and $\varphi_{cl}:\widehat{M}_t \rightarrow R^d$ is the solution of the
 boundary
value problem
\begin{eqnarray}
 \left( {\partial}^2_0 + {\partial}^2_1 \right) \varphi_{cl} & = & 0
\;\;,\nonumber\\
\varphi_{cl}(0,z^1) &=& \widetilde{\varphi}_i(z^1) \;\;,\nonumber\\
\varphi_{cl}(t,z^1) &=& \widetilde{\varphi}_f(z^1) \;\;,\nonumber\\
\partial_1 \varphi_{cl}(z^0,0) &=& \partial_1 \varphi_{cl}(z^0,1) = 0
\;\;.\nonumber
\end{eqnarray}
The relation (\ref{formula})
is well known for Gaussian integrals (as a formula for determinants
\cite{f}) and is supposed to be valid in
general situation \cite{car}. Since we are not
aware of any proof  in the case of boundary interactions, a simple
derivation of the formula (\ref{formula}) is given in the Appendix B.

Using (\ref{formula}) the transition amplitude between arbitrary states
$|\Psi\rangle,|\Psi'\rangle \in$
${\cal H}(
\mbox{{\boldmath R}}_+\times {\cal E}_L) =_{\widehat{e}}
{\cal H}({\cal C}_L)$
can be rewritten in the following form
\begin{eqnarray*}
\langle \Psi | P | \Psi' \rangle &=&
\int\limits^{\infty}_0 d\alpha_f
\int\limits_{{\cal E}_L} {\cal D}^{\alpha_f \hat{e} } \widetilde{x}_f \;
\int\limits^{\infty}_0 d\alpha_i
\int\limits_{{\cal E}_L} {\cal D}^{\alpha_i \hat{e} } \widetilde{x}_i \;
\overline{\Psi[\alpha_f,\widetilde{x}_f]} P[\alpha_f,\widetilde{x}_f;
\alpha_i,\widetilde{x}_i] \Psi'[\alpha_i,\widetilde{x}_i]  \\
&=&
\int\limits_{{\cal W}_L} {\cal D}^{\widehat{e}}\widetilde{\varphi}_f
\int\limits_{{\cal E}_L} {\cal D}^{\alpha_f \widehat{e} }
\widetilde{x}_f \;
\int\limits_{{\cal W}_L} {\cal D}^{\widehat{e}}\widetilde{\varphi}_i
\int\limits_{{\cal E}_L} {\cal D}^{\alpha_i \widehat{e} }
\widetilde{x}_i \;\\
&\times&
\overline{\Psi[\alpha_f,\widetilde{x}_f]}
\ {\rm e}^{ \frac{7-d}{ 32}(\widetilde{\varphi}_f(0) +
\widetilde{\varphi}_f(1))}\\
&\times&
\langle \widetilde{\varphi}_f | \otimes
\langle \widetilde{x}_f\circ \gamma[\widetilde{\varphi}_f]|
\int\limits_0^{\infty}dt \; {\eta (t)}
{\rm e}^{-t(H_0^x + H^{\varphi}_0)}
| \widetilde{\varphi}_i \rangle \otimes
| \widetilde{x}_i\circ \gamma [ \widetilde{\varphi}_i] \rangle \\
&\times&
\Psi'[\alpha_i,\widetilde{x}_i]
{\rm e}^{ \frac{7-d}{ 32}( \widetilde{\varphi}_i(0) +
\widetilde{\varphi}_i(1))}
\;\;\;,
\end{eqnarray*}
where
$\alpha_i =  \int {\rm e}^{ \frac{1}{2} \widetilde{\varphi}_i},
(i\rightarrow f)$.

\noindent In the formula above $H_0^x + H_0^{\varphi}$ is regarded as an
operator
on the space ${\cal H}({\cal W}_L \times {\cal E}_L)$ with the $x$-part
defined by (\ref{ho}) and with the $\varphi$-part given by
$$
H_0^{\varphi} \equiv {\pi\over 2} \left[
 {1\over 2M_{\varphi}} \pi^2_0 +  \sum\limits_{k=1}^{\infty} \left(
{1\over M_{\varphi}} \pi^2_k + M_{\varphi} k^2 {\varphi}^2_k \right)
\right]
\ \ \ \ \ \ \ \ M_{\varphi} = {25 - d \over 96}\;\;;
$$
where
\begin{eqnarray}
\pi(s) &\equiv & -i\frac{\delta}{\delta \widetilde{\varphi}(s)}
= \pi_{0} + 2\sum\limits_{k = 1}^{\infty} \pi_{ k} \cos \pi ks \;\;\;,
\nonumber\\
\widetilde{\varphi}(s) & =
& \varphi_0 + \sum\limits_{k=1}^{\infty} \varphi_k \cos \pi ks\;\;\;.
\label{modes}
\end{eqnarray}

Changing variables
$$
\widetilde{x}_i \longrightarrow \widetilde{x}_i \circ
\gamma[\widetilde{\varphi}_i]^{-1}
\;\;\;\;,\;\;\;(i\rightarrow f)\;\;\;,
$$
and using the relations (\ref{relations}) one gets the  representation
required
\begin{equation}
\langle \Psi |P|\Psi'\rangle = \langle \widetilde{\Psi} |
\int\limits_0^{\infty}dt \; {\eta (t)}
{\rm e}^{-t(H_0^x + H_0^{\varphi})}
| \widetilde{\Psi}'\rangle\;\;\;,
\label{npropag}
\end{equation}
where for each state
$|\Psi\rangle \in {\cal H}({\bf R}_+\times{\cal E}_L)$
the state $|\widetilde{\Psi}\rangle \in {\cal H}({\cal W}_L \times
{\cal E}_L)$ is given by the wave functional
\begin{equation}
\widetilde{\Psi}[\widetilde{\varphi},\widetilde{x}] \equiv
{\rm e}^{ \frac{7+d}{ 32}(\widetilde{\varphi}(0) + \widetilde{\varphi}(1))}
\Psi[{\textstyle \int {\rm e}^{ \frac{1}{2} \widetilde{\varphi}}},
\widetilde{x}\circ \gamma[\widetilde{\varphi}]^{-1}] \;\;\;.
\label{nphyst}
\end{equation}
and the  scalar product
\begin{equation}
\langle \Psi | \Psi' \rangle =
\int\limits_{{\cal W}_L} {\cal D}^{\widehat{e}}\widetilde{\varphi}
\int\limits_{{\cal E}_L} {\cal D}^{ \widehat{e} } \widetilde{x} \;
\overline{\Psi[\widetilde{\varphi},\widetilde{x}]}
 \Psi'[\widetilde{\varphi},
\widetilde{x}]
\;\;\;
\label{gggproduct}
\end{equation}
is used on the r.h.s of (\ref{npropag}).

\subsection{Physical state conditions}

Due to the simple representation (\ref{npropag}) of the transition
amplitude it is convenient to
analyse the physical state conditions in the extended space
${\cal H}({\cal W}_L \times {\cal E}_L)$. We start with the discussion
of the conditions related to the extension itself. Their role is to
select in the extended space the image of the original space of
states under the extension map
$$
{\rm Ext}: {\cal H}({\bf R}_+ \times {\cal E}_L) \ni \Psi
\longrightarrow \widetilde{\Psi} \in
{\cal H}({\cal W}_L \times {\cal E}_L)
\;\;\;,
$$
where $\widetilde{\Psi}$ is given by the formula (\ref{nphyst}).
Using the equations (\ref{difeq}) one can
show that
the functionals  $\Phi[\widetilde{\varphi},\widetilde{x}] \in
{\cal H}({\cal W}_L \times {\cal E}_L)$ of the form (\ref{nphyst})
can be uniquely characterized as functionals
invariant with respect to the following ${\cal D}_L$ action
\begin{equation}
{\cal H}({\cal W}_L \times {\cal E}_L) \ni
\Phi [\widetilde{\varphi},\widetilde{x}]
{ \begin{picture}(55,0)(0,0) \put(5,3){\vector(1,0){45}}
\put(13,10){$\scriptstyle
\gamma \in  {\cal D}_L
$ }
\end{picture}}
\widetilde{\rho}[\gamma]
\Phi [\widetilde{\varphi} \circ \gamma + 2\log \gamma',
\widetilde{x} \circ \gamma]
\in {\cal H}({\cal W}_L \times {\cal E}_L)\;\;\;,
\label{ntrans}
\end{equation}
where
\begin{equation}
\widetilde{\rho}[\gamma]
\equiv {\rm e}^{-{7+d\over 16}(\log \gamma'(0) + \log \gamma'(1) )}
\;\;\;,\;\;\;\widetilde{\rho}[\gamma \circ \delta]
 = \widetilde{\rho}[\gamma]\widetilde{\rho}[\delta]\;\;\;;
\;\;\;
\gamma,\delta \in {\cal D}_L \;\;\;.
\label{nrho}
\end{equation}

Within the old covariant approach the ${\cal D}_L$-action
(\ref{ntrans}) has to be modified in order to meet the
requirement of unitary realization of the residual symmetry.
 The analysis of the transformation
properties of the scalar product (\ref{gggproduct})
with respect to the transformations
$$
{\cal W}_L \times {\cal E}_L \ni
(\widetilde{\varphi},\widetilde{x})
{ \begin{picture}(55,0)(0,0) \put(5,3){\vector(1,0){45}}
\put(13,10){$\scriptstyle
\gamma \in  {\cal D}_L
$ }
\end{picture}}
(\widetilde{\varphi} \circ \gamma + 2\log \gamma',
\widetilde{x} \circ \gamma)
\in {\cal W}_L \times {\cal E}_L\;\;\;,
$$
leads to the following unitary ${\cal D}_L$-action
\begin{equation}
{\cal H}({\cal W}_L \times {\cal E}_L) \ni
\Phi [\widetilde{\varphi},\widetilde{x}]
{ \begin{picture}(55,0)(0,0) \put(5,3){\vector(1,0){45}}
\put(13,10){$\scriptstyle
\gamma \in  {\cal D}_L
$ }
\end{picture}}
\rho[\gamma]
\Phi [\widetilde{\varphi} \circ \gamma + 2\log \gamma',
\widetilde{x} \circ \gamma]
\in {\cal H}({\cal W}_L \times {\cal E}_L)\;\;\;,
\label{nnntrans}
\end{equation}
with
\begin{equation}
\rho[\gamma]
\equiv {\rm e}^{-{1+d\over 16}(\log \gamma'(0) + \log \gamma'(1) )}
\;\;\;,\;\;\;\rho[\gamma \circ \delta]
 =\rho[\gamma]\rho[\delta]\;\;\;;
\;\;\;
\gamma,\delta \in {\cal D}_L \;\;\;.
\label{nnrho}
\end{equation}
The discrepancy between $\widetilde{\rho}$ (\ref{nrho}) derived
from the representation (\ref{npropag}) and $\rho$ (\ref{nnrho})
obtained from the unitarity requirement is related to the fact
that in the old covariant approach one disregards the ghost sector.
The detailed discussion of this point requires the full BRST formulation
which is  beyond the scope of the present paper.
Let us only mention that  $\widetilde{\rho}$ given by the formula
(\ref{nrho}) leads to a unitary realization of the ${\cal D}_L$-symmetry
in the BRST extended space.

The generators of the representation (\ref{nnntrans}) take the form
\begin{eqnarray}
V_k \; \equiv & -& \!\!\!i\int\limits_0^1 ds \; \sin \pi ks
(\widetilde{x}^{\mu})'(s)
\frac{\delta}{\delta \widetilde{x}^{\mu}(s) }
\nonumber \\
&-&\!\!\!i\int\limits_0^1 ds \; \sin \pi ks
(\widetilde{\varphi})'(s)
\frac{\delta}{\delta \widetilde{\varphi}(s) }
- i2\pi k \int\limits_0^1 ds \; \cos \pi ks
\frac{\delta}{\delta \widetilde{\varphi}(s) }
\nonumber \\
&-&\!\!\! i \int\limits_0^1 ds \sin \pi ks
{\frac{\delta}{\delta \gamma(s)}
\rho[\gamma] }_{|\gamma = {\scriptstyle \rm id}_L }
\;\;\;,\;\;\;k\geq 1
\nonumber\\
 = & & \!\!\!\!\!\! V^x_k + V^{\varphi}_k
  \;\;\;;\nonumber
\end{eqnarray}
where $V^x_k$ is given by (\ref{vk}) and
\begin{eqnarray}
V_k^{\varphi}
& = & - {\pi\over 2}\sum\limits_{n = 1}^k  n \varphi_n \pi_{ (k-n)}
+     {\pi \over 2} \sum\limits_{n=1}^{\infty}
\left( n \varphi_n\pi_{n+k} - (n+k)\varphi_{n+k}\pi_{ n}
\right) \nonumber\\
&& + 2\pi k\pi_k  + i\pi {1\over 8}p(k)k\;\;\;,\;\;\;k\geq 1\;\;\;.
\nonumber
\end{eqnarray}
Note that $V_k$ are hermitian with respect to the scalar product
(\ref{gggproduct}) and normally ordered.
The subspace
${\cal H}_{\scriptstyle \rm inv}({\cal W}_L \times {\cal E}_L)$
of states invariant with respect to the action (\ref{nnntrans})
is determined by the equations
$$
V_k |\Phi\rangle = 0\;\;\;,\;\;\;k\geq 1\;\;.
$$

The rest of the physical state conditions can be derived using the method
discussed in Subsect.2.4. The subspace
of $ {\cal H}_{\scriptstyle \rm inv}({\cal W}_L \times {\cal E}_L)$
on which
$$
\int\limits_0^{\infty}dt \; {\eta (t)}
{\rm e}^{-t(H_0^x + H_0^{\varphi})}
$$
reduces to a well defined operator is given by the equations
\begin{equation}
H_k |\Psi\rangle = 0\;\;\;,
\;\;\;V_k  |\Psi\rangle = 0\;\;\;;k=1,...\;\;\;,
\label{nphsc}
\end{equation}
where
$$
H_k \equiv -{i\over \pi k} [H_0^x + H^{\varphi}_0,V_k]
\;=\; H^x_k + H^{\varphi}_k
$$
and
\begin{eqnarray}
H_k^{\varphi}& \equiv& -{i\over \pi k} [H_0^{\varphi},V_k^{\varphi}]
\nonumber\\
&=& { \pi\over 2} \left[ {1\over 2M_{\varphi}}\pi_k\pi_0 +
{1\over 2}\sum\limits_{n=1}^{k} \left( {1\over M_{\varphi}}
\pi_n  \pi_{k-n}
-M_{\varphi} n(k-n){\varphi}_n  {\varphi}_{k-n} \right) \right.
\nonumber \\
&+& \left. \sum\limits_{n=1}^{\infty} \left( {1\over M_{\varphi}}
 \pi_n \pi_{k+n} +
M_{\varphi} n(k+n){\varphi}_n  {\varphi}_{k+n} \right)
\;+\;4M_{\varphi} k^2 \varphi_k \right]\;\;\;.    \nonumber
\end{eqnarray}

The analysis of the integrability conditions of the equations
(\ref{nphsc}) can be simplified by introducing
$$
L_{\pm k} \equiv {1\over \pi} (H_k \pm i
V_k)\;=\; L^x_{\pm k} +L^{\varphi}_{\pm k} \;\;\;,
 \;\;\;k=1,...
$$
where the operators $L^x_k$ are given by (\ref{lxk}) and
$L^{\varphi}_k$ are just the Virasoro generators in
the FCT representation \cite{ct,t} with the central charge $c = 26-d$
\begin{eqnarray*}
L^{\varphi}_k &\equiv& {1\over 2} \sum\limits_{-\infty}^{+\infty}
\beta_{-n} \beta_{k+n} +ikQ\beta_k
\;\;\;,\;\;\;k=\pm 1,\pm 2,...\\
Q &\equiv & 2\sqrt{2 M_{\varphi}} = \sqrt{{25 -d \over 12}}\;\;\;,\\
\beta_0^{\mu} &\equiv& {1\over \sqrt{2M_{\varphi}}} \pi_0\;\;\;,\\
\beta_n^{\mu } &\equiv & {1\over \sqrt{2}}
\left( {1\over \sqrt{M_{\varphi}}}\pi_{ n}-
i \sqrt{M_{\varphi}} n \varphi_n \right)
\;\;\;,\;\;\;n=\pm 1,\pm 2,...\;\;\;;
\end{eqnarray*}
$$
\left[\beta_n,\beta_m\right] =
 n \delta_{n,-m}\;\;\;\;\;
,\;\;\;\;\;\beta_n^+ = \beta_{-n}\;\;\;.
$$
 The integrability conditions take the following form
$$
\left[ L_n ,L_m \right] =
(n-m) L_{n+m} +
{26 \over 12}
\delta_{n,-m}(n^3-n)\;\;\;,
$$
where
\begin{eqnarray}
L_0& \equiv & L_0^x + L^{\varphi}_k\;\;\;, \nonumber\\
       L^{\varphi}_0 &\equiv& {1\over 2}\beta_0^2 +
\sum\limits_{n=1}^{\infty} \beta_{-n}\beta_n + \frac{Q^2 }{2}\;\;\;,
\nonumber
\end{eqnarray}
and $L_0^x$ is given by (\ref{lo}).

As in the case of critical string relaxing the strong form of
the equations (\ref{nphsc}) one gets the conditions for
the off-mass-shell physical states
\begin{equation}
L_k|\Psi\rangle = 0\;\;\;,\;\;\;k\geq 1\;\;\;.
\label{noph}
\end{equation}
The structure of the space
${\cal H}_{\scriptstyle \rm ph}^{\scriptstyle \rm off}({\cal W}_L
\times {\cal E}_L)$
of solutions to the equations (\ref{noph}) is similar to that
of critical string theory.
As we shall see in the next subsection, the quotient
space of the off-mass-shell physical states modulo null states
can be uniquely characterized as the space of solutions of (\ref{noph})
satisfying the "quantum" light-cone gauge conditions
$$
\alpha^+_k|\Psi\rangle \equiv \left(i \alpha_k^0 +
\alpha_k^{d-1} \right) |
\Psi\rangle = 0\;\;\;,\;\;\;k\geq 1\;\;\;.
$$
In particular repeating the considerations presented in Subsect.2.4.
one can
derive the on-mass-shell condition which
 in the space ${\cal H}({\cal W}_L\times{\cal E}_L)$
 takes the following form
\begin{equation}
(L_0 -1)|\Psi\rangle = 0\;\;\;.
\label{nmshell}
\end{equation}

\subsection{DDF construction and no-ghost theorem}

In this subsection we present an explicit construction of
physical states of the relativistic theory.
Performing the Wick rotation in the conditions
(\ref{noph}), (\ref{nmshell}) one gets
\begin{equation}
L_k|\Psi\rangle = 0\;\;\;,\;\;\;k\geq 1\;\;\;,
\;\;\;(L_0 -1)|\Psi\rangle = 0\;\;\;,
\label{nph}
\end{equation}
where
\begin{eqnarray*}
L_n&=& \frac{1}{2}\sum_{m=-\infty}^{\infty}\alpha_m\cdot
\alpha_{n-m}
+\frac{1}{2}\sum_{m=-\infty}^{\infty}\beta_m\beta_{n-m}+inQ\beta_n
\ \ \ \ \ n\ne 0\\
L_0&=& \frac{1}{2}\alpha_0\cdot\alpha_0+\frac{1}{2}\beta_0^2+
\sum_{m=1}^{\infty}\alpha_{-m}\cdot \alpha_m
+\sum_{m=1}^{\infty}\beta_{-m}\beta_m+\frac{Q^2}{2}
\end{eqnarray*}
and
$$
\left[\beta_m,\beta_n\right]= m\delta_{m,-n}\;\;\;,\;\;\;
\left[\alpha^{\mu}_m,\alpha^{\nu}_n\right]= m\eta^{\mu\nu}
\delta_{m,-n}\;\;\;,\;\;\;
\eta^{\mu\nu}= diag(-1,+1,..,+1)\;\;\;.
$$

Following the DDF approach \cite{ddf} we introduce
the operators of "position" and "momentum" :
\begin{eqnarray*}
X^{\mu}(\theta)&=& x^{\mu}_0+\alpha^{\mu}_0\theta+
\sum_{k\ge 1} \frac{i}{k}\left(\alpha_k^{\mu}{\rm e}^{-ik\theta}-
\alpha_k^{\mu +}{\rm e}^{ik\theta}\right)\;\;\;,\\
P^{\mu}(\theta)&=& \alpha^{\mu}_0+
\sum_{k\ge 1} \left(\alpha_k^{\mu}{\rm e}^{-ik\theta}+
\alpha_k^{\mu +}{\rm e}^{ik\theta}\right)\;\;\;,\\
\Phi(\theta)&=& \varphi_0+\beta_0\theta+
\sum_{k\ge 1} \frac{i}{k}\left(\beta_k{\rm e}^{-ik\theta}-
\beta_k^{+}{\rm e}^{ik\theta}\right)\;\;\;,\\
\Pi(\theta)&=& \beta_0+
\sum_{k\ge 1} \left(\beta_k{\rm e}^{-ik\theta}+
\beta_k^{+}{\rm e}^{ik\theta}\right)       \;\;\;.
\end{eqnarray*}
Using the relations
\begin{eqnarray*}
\left[L_n,\alpha^{\mu}_m\right]&=& -m\alpha^{\mu}_{m+n}\;\;\;,\\
\left[L_n,\beta_m\right]&=& -m\beta_{m+n}+iQn^2\delta_{n,-m}\;\;\;,\\
\left[L_n,L_m\right]&=& (n-m)L_{n+m}+\frac{26}{12}(n^3-n)\delta_{n,-m}
\;\;\;,
\end{eqnarray*}
one gets
\begin{eqnarray}
\left[X^{\mu}(\theta),X^{\nu}(\theta')\right]&=&
-i\pi\eta^{\mu\nu}{\rm sgn}
(\theta-\theta')\;\;\;,\nonumber\\
\left[P^{\mu}(\theta),P^{\nu}(\theta')
\right]&=& 2i\pi\eta^{\mu\nu}\delta'(\theta-
\theta')\;\;\;,\nonumber\\
\left[L_n,P^{\mu}(\theta)\right]&=& -i\frac{d}{d\theta}\left(P^{\mu}
{\rm e}^{in\theta}\right)\;\;\;,\nonumber\\
\left[L_n,\Pi(\theta)\right]&=&
-i\frac{d}{d\theta}\left(\Pi^{\mu}
{\rm e}^{in\theta}\right)
+iQn^2 {\rm e}^{in\theta}               \;\;\;.
\label{nonhom}
\end{eqnarray}

Let us consider the state $|p_L,p^{\mu}\rangle $ satisfying
\begin{eqnarray}
\alpha_n^{\mu}\vert p_L,p^{\mu}\rangle &=& \delta_{n0}
  p^{\mu}\vert p_L,p^{\mu}\rangle \;\;\;,\nonumber\\
\beta_n \vert p_L,p^{\mu}\rangle &=& \delta_{n0}
p_L \vert p_L,p^{\mu}\rangle\;\;\;,
\label{vacuum}\\
(L_0 -1)|p_L,p^{\mu}\rangle&=&0\;\;\;.\nonumber
\end{eqnarray}
For $p^{\mu} \ne 0$ there exists a vector $k$
such that $k^{\mu}k_{\mu}=0$ and $k^{\mu}p_{\mu}=1$.

The construction of vertex operators, which acting on the state
(\ref{vacuum})
generate positive norm physical states with transverse
excitations is the same as in the case of Nambu-Goto string.
One gets the operators
$$
A^i_n= \frac{1}{2\pi}
\int^{2\pi}_0 d\theta\ P^i(\theta){\rm e}^{ink\cdot X(\theta)}
\;\;\;,\;\;\; i=1,...,d-2\;\;,\;\;n\geq 1\;\;,
$$
satisfying the relations
\begin{eqnarray}
\left[A^i_m,A^j_n\right]&=& m\delta^{ij}\delta_{m,-n}\;\;\;,\nonumber\\
\left[L_k,A^i_m\right]&=& 0\;\;\;,\;\;\;k\geq 0\label{altr}\\
A^{i+}_n&=& A^i_{-n}\;\;\;.\nonumber
\end{eqnarray}
Due to the $n^2$-term in the commutation relation
(\ref{nonhom}) the construction of the vertex operator generating states
with excitation in the Liouville direction is slightly more complicated.
In order to compensate this term one can use a modification
 introduced
by Brower in his construction of the vertex generating longitudinal
excitations in the Nambu-Goto string \cite{b} and write
$$
A^L_n= \frac{1}{2\pi}
\int^{2\pi}_0 d\theta\ \left(\Pi(\theta)-Q(k\cdot\dot P(\theta))
(k\cdot P(\theta))^{-1}\right){\rm e}^{ink\cdot X(\theta)}\;\;\;,\;\;\;
n\geq 1\;\;\;.
$$
In contrast to Brower's longitudinal vertex the operator
above satisfies the  relations analogous to $A_n^i$:
\begin{eqnarray}
\left[A^L_m,A^L_n\right]&=&m\delta_{m,-n}\;\;\;,\nonumber\\
\left[L_k,A^L_m\right]&=& 0\;\;\;,\;\;\;k\geq 0\;\;\;,
\label{allo}\\
\left[A^i_m,A^L_n\right]&=& 0 \;\;\;,\nonumber\\
A^{L+}_n&=&A^L_{-n}\;\;\;.\nonumber
\end{eqnarray}
All the states generated by the operators $A^i_n, A^L_m$ from
the states $|p_L,p^{\mu}\rangle$ with $p^{\mu}\ne 0$ we call
the DDF states.
The commutation relations (\ref{altr}),(\ref{allo}) imply
that the DDF states  are physical
states with positive norm. The inverse statement can be formulated
as follows

{\bf Theorem.} {\it Any solution of the equations {\rm (\ref{nph})}
in the Hilbert space ${\cal H}({\cal W}_L\times {\cal E}_L)$
is of the form
$$
|\Phi\rangle = |\Psi\rangle + |ns\rangle\;\;\;,
$$
where $|\Psi\rangle$ is either a DDF state or one of the states
\begin{equation}
\vert p_L=\pm  Q',p^{\mu}=0\rangle
\;\;\;,\;\;\; Q'=\sqrt{\frac{d-1}{12}}\;\;\;,
\label{ovacuum}
\end{equation}
and $|ns\rangle$ is a null spurious state, i.e. $|ns\rangle$ is
orthogonal to all physical states and is a linear combination
of states of the form $L_{-n}|\chi\rangle , n\geq 1$.}

A counterpart of the theorem above for the standard free field
realization of the Virasoro algebra with the central charge
$c=26$ and $\alpha_0 =1$ has been proved long time ago by
Goddard and Thorn \cite{gt}. Since the first of the two proofs given
in \cite{gt} is based only on the algebraic properties of the
DDF operators it  applies without modification in the
present case. In fact for a fixed kinematical configuration
given by a state (\ref{vacuum}) with $p^{\mu}\ne 0$ one
can introduce the operators
$$
K_n=k\cdot \alpha_n
$$
satisfying the algebra
$$
\left[K_m,K_n\right]=0\ \ \ \left[L_m,K_n\right]=-nK_{n+m}
\ \ \ K_n^+=K_{-n}
$$
The algebra of $A$'s, $K$'s and $L$'s is exactly the same as in
\cite{gt}. The only difference is the number of
positive-norm directions ($d-1$ instead of $24$) which however does not
alter the reasoning given in \cite{gt}.
To complete the proof of the present version of the theorem
let us observe that in the space
${\cal H}({\cal W}_L\times {\cal E}_L)$
the operator $\beta_0$ is self-adjoint and therefore has a real spectrum.
Consequently the only physical states with all components of the spacetime
momenta equal zero are the lowest states given by (\ref{ovacuum}) and
all excited positive norm physical states can be achieved by the DDF
construction.

As a simple consequence of the theorem above
and the algebra (\ref{altr}), (\ref{allo}), one gets
the no-ghost theorem for the model given by the equations (\ref{nph}).
The physical content of the model can be easily inferred
using the DDF construction. One can
show that the positive norm physical states could be uniquely
characterized
in terms of the space-time spin and momenta along with an additional
internal quantum number represented by the operator $\beta_0$.
For each particular eigenvalue $p_L$ of $\beta_0$ the physical states
satisfying $\beta_0|\psi\rangle = p_L|\Psi\rangle$ form the Hilbert
space ${\cal H}(p_L)$ describing a free noncritical string with the
intercept $\alpha_0 = {d-1\over 24}- {p_L^2\over 2}$.
For $p_L^2 < {d-1\over 24}$ the lowest states in the space
${\cal H}(p_L)$ are tachyons.

Let us note that
the  theory described by the Hilbert space ${\cal H}(p_L)$
differs from that with the same intercept and
obtained by the dimension reduction from the critical Nambu-Goto
string \cite{sch}.
In fact, if $T^d(N)$ is the number of states on the level $N$
generated by the d-component oscillators, then the numbers of the
positive norm physical states on the level $N$  is $T^{d-1}(N)$
in the Polyakov string while in the reduced Nambu-Goto string
one gets  $T^{d-1}(N)-T^{d-1}(N-1)$ \cite{b}.
\vspace{7mm}

\section{Conclusions}

The main result of the present paper
is that the Polyakov path integral
over surfaces does lead to a  free quantum theory of 1-dim
extended relativistic system in the range $1<d<25$.
The resulting theory is equivalent to the
FCT  "massive" string model.
As far as the free theory  is concerned
this model can be directly compared with the noncritical
Polyakov dual model in the range $1<d<25$.
In the commonly used radial gauge the "massive" string is given
by the realization (\ref{bcharge}),(\ref{herm}) of the Virasoro
algebra. From this point of view the FCT string can be regarded as
a special version of the 2-dim Liouville model coupled to d-copies
of the free scalar conformal field theory , characterized by the
equations
\begin{eqnarray}
\mu& =&0 \;\;\;,
\label{mu}
\\
(\beta_0 + iQ)^+& =& \beta_0 + iQ \;\;\;.
\label{her}
\end{eqnarray}
Within the Polyakov dual model approach the equations (\ref{mu},
\ref{her}) are just a special (in a sense trivial) choices of
free "parameters": the cosmological constant and the scalar
product in the space of states. In the present approach
the first equation is an assumption partly justified by the
requirement of stability while the second one
uniquely follows from the interpretation of the model as
a quantum mechanics of 1-dim extended  relativistic system.
Since the equations are crucial for the physical interpretation of the
model we shall briefly discuss their origin within the present approach.

First of all let us stress that
our whole derivation is based on the particular choice
of boundary conditions for the string trajectories.
In the "matter" sector these boundary conditions have been
first introduced in \cite{dop}. More recently it was shown that
they are relevant for constructing the
off-shell-critical string amplitudes \cite{jas}. The derivation of the
boundary conditions in the metric sector, based on the geometry
of the space of trajectories has been presented in
\cite{ja}. The outcome of this analysis is
that as far as the interpretation of the
Polyakov path integral as a sum over bordered surfaces is assumed
the boundary conditions in this sector are uniquely determined.
Finally the relevance of these boundary conditions in the
noncritical Polyakov string theory has been confirmed in our
previous paper \cite{jm} concerning the quasiclassical
calculation of
the static potential in the range $1<d<25$. All these results
along with the considerations of Sect.2 show that within
the Feynman functional quantization scheme in the $(M,n)$-gauge
the choice of boundary conditions is unique.

The second important point in our paper is the assumption
concerning the vanishing cosmological constant.
As it was discussed in Subsect.3.2 in the case of rectangle
and for $\mu \neq 0$ there is no classical solutions of
the Liouville equation of motion in the space of conformal
factors over which one has to integrate in the path integral
representation of the transition amplitude. If we interpret
the absence of classical extremum as an indication of instability
of the system the only consistent choice is $\mu = 0$.
Whether or not this conclusion is fully justified is still
an open problem. One possible approach is to assume
that the generalized Forman formula still holds in the case of
the bulk exponential interaction (which can be justified to some extent
by means of the perturbation expansion \cite{car})
and then to analyse the
resulting theory in the extended space.
Whatever the final understanding of the model with
$\mu > 0$  would be, the
simplest case $\mu = 0$ yields a consistent free theory and
it is a nontrivial and interesting problem  to investigate
 the "joining-splitting"
interaction in this model.

While the equation (\ref{mu}) is an assumption more or less justified
by our choice of boundary conditions, the second equation (\ref{her})
uniquely follows from them. In fact the central technical point
of our approach -- the generalized Forman formula - yields not
only the simple expression for the transition amplitude in the
extended space but also the inner product in the conformal factor
sector.

The relation of the FCT realization of the Virasoro
algebra with the Polyakov path integral over surfaces has been
known since the first attempts to quantize the Liouville
theory \cite{cct}. More recently it became a standard tool in
analysing the physical states \cite{pb} and correlators
\cite{dof,corel} in the
Liouville gravity coupled to the conformal matter.
The question arises what we have learned from lengthy
derivation of the particular realization
characterized by the equation (\ref{her}).

The most important lesson is the path integral formulation of
the FCT string. As it was emphasized in the
introduction it paves a way for analysing the
"joining-splitting" interaction in this string model. Note
that the lack of the path integral formulation was a basic
obstacle for developing the interacting theory of the "massive"
string twenty years ago \cite{ct}.

A related issue is the target space interpretation of the free Polyakov
model in the range $1<d<25$.
As far as this interpretation is concerned there are no
physical states with imaginary (or complex) Liouville momentum
in the model.
This is in contrast with the
Polyakov noncritical dual model interpreted as the 2-dim Euclidean
gravity coupled to the conformal matter.
This interpretation concentrates on the word sheet
physics bringing all the questions of the theory of 2-dim statistical
systems. In particular one of the prominent observables in
these framework is the area operator getting complex
for the central charge of the matter sector in the range
$1<d<25$. The physical states with the imaginary Liouville
momenta are therefore indispensable within this interpretation  and
lead to the "unstable" critical
behaviour \cite{ddk,sei}.

The derivation of the FCT free string from
the Polyakov path integral over bordered surfaces given in Sect.3
sheds new light on the role the conformal factor plays in
the Polyakov string model.
The noncritical string model we have started with was originally
described in terms of  the
variables $\{ \alpha, x^{\mu}(\sigma) \}_{\sigma \in [0,1]}$.
Roughly speaking the dynamics is given by an on-mass-shell
condition (the string wave equation) and the kinematical
requirement. It means that in a fixed frame in the Minkowski target space
all nonzero modes of the $x^+$ variable are
unphysical and there is a relation for the momenta conjugate
to the zero modes $\{x^{\mu}_0\}_{\mu=0,...,d-1}$.
It follows that the set $\{\alpha,x^{\mu}_0,
x^-_k,x^i_k\}_{k=1,...}^{\mu =0,...,d;i=1,...,d-1}$
is a complete system of commuting physical micro-observables.
We have used the prefix micro- in order to distinguish them
from the "true" physical macro-observables which are given
by the generators of the Poincare group in the Minkowski
target space.
Actually the spectra of the macro-observables are of the
main interest in the free theory as they provide a relativistic
particle interpretation of the string physical states.

In terms of the physical micro-observables
$\{\alpha,x^{\mu}_0,
x^-_k,x^i_k\}_{k=1,...}^{\mu =0,...,d;i=1,...,d-1}$
the geometrical
content of the model is clear - it is a theory of the free
parametrized string with internal length.
In this formulation however
the macro-observables are not diagonal. Moreover
we do not know how to derive the on-mass-shell
condition in these variables from the complicated
form of the transition amplitude. The idea of extension
we applied to deal with this problem
was to introduce an auxiliary variable $\varphi(\sigma)$
along with some constraints ensuring the equivalence
with the original theory. This allows for expressing
the original set of physical micro-observables in terms
of a new one $\left\{\varphi_0,x^{\mu}_0,\varphi_k,x^i_k
\right\}^{\mu =0,...,d;i=1,...,d-1}_{k=1,...}$. As follows from the
DDF construction given in Subsect.3.4 in the new variables
the macro-observables are diagonal and the relativistic
particle content of the model can be easily inferred.
The role of the conformal factor is therefore to express in a
convenient way
the influence of the nontrivial dynamics of
the internal length $\alpha$ and
the longitudinal excitations $\{x^-_k\}_{k\geq1}$
on the particle spectrum of the noncritical
relativistic string model. In this sense
the Liouville theory describes the dynamics of the longitudinal
modes.

Although the FCT model satisfies all the
consistency conditions of formal relativistic quantum mechanics
its physical content  is not quite
satisfactory. First of all it contains an  internal
quantum number entering the on-mass-shell condition which entails
an undesirable continuous range of intercepts. Secondly for some
values of this quantum number one gets tachyonic states on the
lowest level.

In the free string theory the zero mode of
the "Liouville momenta"  is conserved and the theory
can be truncated at any real value of $p_L$. On the other
hand the appearance of this additional internal degree of freedom
is a consequence of our choice of boundary conditions for
string trajectories involving the internal string length $\alpha$.
The relation between $\alpha$ and $p_L$ is a part of  the
relation  between two sets of physical micro-observables discussed
above
\begin{equation}
\left\{\alpha,x^{\mu}_0,
x^-_k,x^i_k\right\}_{k=1,...}^{\mu =0,...,d;i=1,...,d-1}
\leftarrow \! \! \rightarrow
\left\{\varphi_0,x^{\mu}_0,\varphi_k,x^i_k
\right\}^{\mu =0,...,d;i=1,...,d-1}_{k=1,...}\;\;\;.
\label{rela}
\end{equation}

Using the constraint equations one can easily express the
set of micro-observables
$\left\{\alpha,x^{\mu}_0,
x^-_k,x^i_k\right\}_{k=1,...}^{\mu =0,...,d;i=1,...,d-1}$
in terms of
$\left\{\varphi_0,x^{\mu}_0,\varphi_k,x^i_k
\right\}^{\mu =0,...,d;i=1,...,d-1}_{k=1,...}$
The opposite relation is however very complicated
and we have not found any convincing method of removing the
$\alpha$-dependence within the first quantized theory.
It seems that the truncation
is essentially the problem of the interacting theory where $\alpha$
plays the role similar to that of the "length" parameter in
the "covariantized" light cone formulation of the critical string
field theory \cite{hata}. Let us only mention that the results concerning
noncritical Polyakov string with fixed ends \cite{jm} and the naive
consideration of the "joining-splitting" interaction suggest
a consistent truncation at $p_L = 0$.

The second problem with the physical interpretation of the
FCT massive string is the presence of tachyons in its spectrum.
Since the structure of the model is similar to that of
the critical string, one may expect that the problem  can
be solved in the fermionic noncritical Polyakov string by a counterpart
of GSO projection
\cite{gso}. The crucial issue here is an appropriate choice
of boundary conditions for the fermionic string trajectories.
Since the geometry of the fermionic path integral is far
less understood than that of the bosonic one (the infinite-dimensional
supergeometry virtually does not exist
\cite{nb}) the methods we have
used to determine the boundary conditions in the bosonic case
are not available.
This makes the problem of a "super" generalization of our
approach more difficult than the construction of the
supersymmetric noncritical Polyakov dual model \cite{super}.

The considerations of the present paper are entirely devoted
to the so called "old" covariant formulation of the free open bosonic
string. This leaves a number of interesting questions concerning
the first quantized theory.

{\it Closed string}.
In contrast to the open string there is no 1-dim conformal
anomaly and the natural scalar product is diff-invariant. The
issue of the normal ordering of the Virasoro generators appears
if one tries to separate the left and right movers. The  additional
complication is that this decomposition
is not invariant with respect to the residual $S^1$-symmetry.
For these reasons the free closed noncritical string
is an interesting problem and would lead
to a better insight into the relation between the geometry of
the scalar product and the ordering problem.

{\it BRST formulation}. Within the path integral approach of this paper
the idea is in a sense opposite to that of the "old" covariant
formulation. One starts with the same path integral representation
of the transition amplitude in the extended space of boundary
conditions. However instead of restricting oneself to a subspace of states
on which the corresponding operator is well defined and invertible,
we are looking for yet another (BRST) extension which allows to
represent the transition amplitude as
a special matrix element of an invertible operator.
In the case of the critical string this
idea can be easily realized \cite{jask} leading to the well known
covariant BRST formulation.
In the case of noncritical string the problem gets complicated
due to nontrivial coupling of the conformal factor
to the zero modes of the ghost sector. In particular the two extension
procedures do not commute. Note that the complete
BRST formulation of the free theory is interesting at least
for two reasons. First it should provide a clarification of
the discrepancy between the realization of the residual
${\cal D}_L$-symmetry calculated from the extension procedure and
the unitary one in the "old" covariant approach. Secondly,
it paves the way for the field theory formulation which
gives new tools for investigating the interacting
theory. In particular the problem of the joining-splitting
interaction vertex can be posed as
the problem of the BRST-invariant extension
of the corresponding functional delta function in the variables
$\{ \alpha, x^{\mu}(\sigma) \}_{\sigma \in [0,1]}$. Note that
the role of $\alpha$  in the interacting theory is
especially clear in this formulation. The values of $\alpha$
determine the
way in which two parametrized strings form a third one.

{\it Light-cone formulation.} The idea of this approach is to
explicitly implement the basic kinematical requirement of
the relativistic quantum mechanic of 1-dim extended objects by
constructing the transition amplitude in a fixed reference
system as a sum over causal string trajectories in the Minkowski
target space. The main difficulty with respect to the critical string
consists in the fact that the constraints appear in the process
of quantization and one cannot describe relevant string trajectories
in terms of
"true" classical variables. A related problem is an appropriate
choice of the space of internal metrics corresponding to
causal string trajectories in the Minkowski target space.
The form of the DDF-states given
in Subsect.3.4 suggest however that this formulation has the
structure very similar to that of the critical string theory.

{\it Operator - states correspondence.} The basic tool in
calculating the on-shell critical string amplitudes is
the so called operator formalism based on the
possibility of reproducing string wave functionals
corresponding to the physical states by functional
integral over half-disc with a local operator insertion.
In the present case this equivalence still holds, as
can be inferred from the construction of DDF states.
The operator-state correspondence along with the solution
of the previous problem form basic ingredients of the
Mandelstam method \cite{m} of constructing
the on-mass-shell "massive" string amplitudes.

As it was emphasized in the introduction the most interesting
open question is whether the "joining-splitting" interaction leads
to a consistent interacting theory of the FCT string.
We have already mentioned two possible approaches to solve this
problem: the BRST and the light-cone formulations.
We conclude this section by a brief discussion of
only one aspect of the interacting theory which is to some
extent independent of a particular formulation.

The path integral formulation of the free FCT string
derived in this paper
involves the assumption that the cosmological constant vanishes.
The question arises whether this assumption can be maintained also
in the interacting theory.
To analyse this problem let us consider the three point
off-shell noncritical string amplitude. As in the case of the critical
string theory \cite{jas} it is given by the path
integral over  string trajectories connecting three
prescribed string configurations. On the tree level
one has to sum over trajectories of the topology of hexagon.
The construction of the path integral representation of
the corresponding off-shell amplitude is exactly the same as
in the case of rectangle-like trajectories relevant for the
string propagator. Choosing the hyperbolic hexagon as
a model manifold one gets in the conformal gauge an expression
involving the path integral over all conformal factors satisfying
the homogeneous Neumann boundary conditions. The reasoning
we have used in the case of rectangle to derive the condition $\mu=0$
now leads to a positive cosmological constant. This makes the
resulting path integral prohibitively complicated.
Up to now the only method to deal with the resulting theory is to
impose the requirement of the conformal invariance and then
to translate the problem into the language of 2-dim conformal
field theory. This leads however to the complex Liouville action
-- yet another manifestation of the $c=1$ barrier.

\begin{picture}(400,200)(30,0)
\thicklines
\put(155,40){\makebox(20,10)[t]{(a)}}
\put(290,40){\makebox(20,10)[t]{(b)}}
\put(220,10){\makebox(20,10){Fig.1.} }
\put(260,110){\line(1,0){40}}
\put(260,110){\line(0,1){60}}
\put(260,170){\line(1,0){90}}
\put(350,170){\line(0,-1){100}}
\put(350,70){\line(-1,0){50}}
\put(300,70){\line(0,1){70}}
\setlength{\unitlength}{0.240900pt}
\ifx\plotpoint\undefined\newsavebox{\plotpoint}\fi
\sbox{\plotpoint}{\rule[-0.500pt]{1.000pt}{1.000pt}}%
\font\gnuplot=cmr10 at 10pt
\gnuplot
\sbox{\plotpoint}{\rule[-0.500pt]{1.000pt}{1.000pt}}%
\put(895,474){\usebox{\plotpoint}}
\put(891,472.92){\rule{0.964pt}{1.000pt}}
\multiput(893.00,471.92)(-2.000,2.000){2}{\rule{0.482pt}{1.000pt}}
\put(887,474.92){\rule{0.964pt}{1.000pt}}
\multiput(889.00,473.92)(-2.000,2.000){2}{\rule{0.482pt}{1.000pt}}
\put(883,476.92){\rule{0.964pt}{1.000pt}}
\multiput(885.00,475.92)(-2.000,2.000){2}{\rule{0.482pt}{1.000pt}}
\put(879,479.42){\rule{0.964pt}{1.000pt}}
\multiput(881.00,477.92)(-2.000,3.000){2}{\rule{0.482pt}{1.000pt}}
\put(876,481.92){\rule{0.723pt}{1.000pt}}
\multiput(877.50,480.92)(-1.500,2.000){2}{\rule{0.361pt}{1.000pt}}
\put(872,484.42){\rule{0.964pt}{1.000pt}}
\multiput(874.00,482.92)(-2.000,3.000){2}{\rule{0.482pt}{1.000pt}}
\put(868,486.92){\rule{0.964pt}{1.000pt}}
\multiput(870.00,485.92)(-2.000,2.000){2}{\rule{0.482pt}{1.000pt}}
\put(865,489.42){\rule{0.723pt}{1.000pt}}
\multiput(866.50,487.92)(-1.500,3.000){2}{\rule{0.361pt}{1.000pt}}
\put(861,492.42){\rule{0.964pt}{1.000pt}}
\multiput(863.00,490.92)(-2.000,3.000){2}{\rule{0.482pt}{1.000pt}}
\put(858,495.42){\rule{0.723pt}{1.000pt}}
\multiput(859.50,493.92)(-1.500,3.000){2}{\rule{0.361pt}{1.000pt}}
\put(854,498.42){\rule{0.964pt}{1.000pt}}
\multiput(856.00,496.92)(-2.000,3.000){2}{\rule{0.482pt}{1.000pt}}
\put(851,501.42){\rule{0.723pt}{1.000pt}}
\multiput(852.50,499.92)(-1.500,3.000){2}{\rule{0.361pt}{1.000pt}}
\put(848,504.42){\rule{0.723pt}{1.000pt}}
\multiput(849.50,502.92)(-1.500,3.000){2}{\rule{0.361pt}{1.000pt}}
\put(844,507.42){\rule{0.964pt}{1.000pt}}
\multiput(846.00,505.92)(-2.000,3.000){2}{\rule{0.482pt}{1.000pt}}
\put(841,510.42){\rule{0.723pt}{1.000pt}}
\multiput(842.50,508.92)(-1.500,3.000){2}{\rule{0.361pt}{1.000pt}}
\put(838,513.42){\rule{0.723pt}{1.000pt}}
\multiput(839.50,511.92)(-1.500,3.000){2}{\rule{0.361pt}{1.000pt}}
\put(835,516.42){\rule{0.723pt}{1.000pt}}
\multiput(836.50,514.92)(-1.500,3.000){2}{\rule{0.361pt}{1.000pt}}
\put(831.42,520){\rule{1.000pt}{0.964pt}}
\multiput(832.92,520.00)(-3.000,2.000){2}{\rule{1.000pt}{0.482pt}}
\put(828.92,524){\rule{1.000pt}{0.723pt}}
\multiput(829.92,524.00)(-2.000,1.500){2}{\rule{1.000pt}{0.361pt}}
\put(827,526.42){\rule{0.723pt}{1.000pt}}
\multiput(828.50,524.92)(-1.500,3.000){2}{\rule{0.361pt}{1.000pt}}
\put(823.42,530){\rule{1.000pt}{0.964pt}}
\multiput(824.92,530.00)(-3.000,2.000){2}{\rule{1.000pt}{0.482pt}}
\put(820.92,534){\rule{1.000pt}{0.964pt}}
\multiput(821.92,534.00)(-2.000,2.000){2}{\rule{1.000pt}{0.482pt}}
\put(819,537.42){\rule{0.723pt}{1.000pt}}
\multiput(820.50,535.92)(-1.500,3.000){2}{\rule{0.361pt}{1.000pt}}
\put(815.92,541){\rule{1.000pt}{0.964pt}}
\multiput(816.92,541.00)(-2.000,2.000){2}{\rule{1.000pt}{0.482pt}}
\put(813.92,545){\rule{1.000pt}{0.723pt}}
\multiput(814.92,545.00)(-2.000,1.500){2}{\rule{1.000pt}{0.361pt}}
\put(811.42,548){\rule{1.000pt}{0.964pt}}
\multiput(812.92,548.00)(-3.000,2.000){2}{\rule{1.000pt}{0.482pt}}
\put(808.92,552){\rule{1.000pt}{0.964pt}}
\multiput(809.92,552.00)(-2.000,2.000){2}{\rule{1.000pt}{0.482pt}}
\put(806.92,556){\rule{1.000pt}{0.964pt}}
\multiput(807.92,556.00)(-2.000,2.000){2}{\rule{1.000pt}{0.482pt}}
\put(804.92,560){\rule{1.000pt}{0.723pt}}
\multiput(805.92,560.00)(-2.000,1.500){2}{\rule{1.000pt}{0.361pt}}
\put(802.92,563){\rule{1.000pt}{0.964pt}}
\multiput(803.92,563.00)(-2.000,2.000){2}{\rule{1.000pt}{0.482pt}}
\put(800.92,567){\rule{1.000pt}{0.964pt}}
\multiput(801.92,567.00)(-2.000,2.000){2}{\rule{1.000pt}{0.482pt}}
\put(799.42,571){\rule{1.000pt}{0.964pt}}
\multiput(799.92,571.00)(-1.000,2.000){2}{\rule{1.000pt}{0.482pt}}
\put(797.92,575){\rule{1.000pt}{0.964pt}}
\multiput(798.92,575.00)(-2.000,2.000){2}{\rule{1.000pt}{0.482pt}}
\put(796.42,579){\rule{1.000pt}{0.964pt}}
\multiput(796.92,579.00)(-1.000,2.000){2}{\rule{1.000pt}{0.482pt}}
\put(794.92,583){\rule{1.000pt}{0.964pt}}
\multiput(795.92,583.00)(-2.000,2.000){2}{\rule{1.000pt}{0.482pt}}
\put(793.42,587){\rule{1.000pt}{0.964pt}}
\multiput(793.92,587.00)(-1.000,2.000){2}{\rule{1.000pt}{0.482pt}}
\put(792.42,591){\rule{1.000pt}{0.964pt}}
\multiput(792.92,591.00)(-1.000,2.000){2}{\rule{1.000pt}{0.482pt}}
\put(790.92,595){\rule{1.000pt}{0.964pt}}
\multiput(791.92,595.00)(-2.000,2.000){2}{\rule{1.000pt}{0.482pt}}
\put(789.42,599){\rule{1.000pt}{1.204pt}}
\multiput(789.92,599.00)(-1.000,2.500){2}{\rule{1.000pt}{0.602pt}}
\put(788.42,604){\rule{1.000pt}{0.964pt}}
\multiput(788.92,604.00)(-1.000,2.000){2}{\rule{1.000pt}{0.482pt}}
\put(787.42,608){\rule{1.000pt}{0.964pt}}
\multiput(787.92,608.00)(-1.000,2.000){2}{\rule{1.000pt}{0.482pt}}
\put(786.42,616){\rule{1.000pt}{0.964pt}}
\multiput(786.92,616.00)(-1.000,2.000){2}{\rule{1.000pt}{0.482pt}}
\put(785.42,620){\rule{1.000pt}{0.964pt}}
\multiput(785.92,620.00)(-1.000,2.000){2}{\rule{1.000pt}{0.482pt}}
\put(789.0,612.0){\usebox{\plotpoint}}
\put(784.42,633){\rule{1.000pt}{0.964pt}}
\multiput(784.92,633.00)(-1.000,2.000){2}{\rule{1.000pt}{0.482pt}}
\put(787.0,624.0){\rule[-0.500pt]{1.000pt}{2.168pt}}
\put(784,644.42){\rule{0.482pt}{1.000pt}}
\multiput(785.00,643.92)(-1.000,1.000){2}{\rule{0.241pt}{1.000pt}}
\put(780,643.92){\rule{0.964pt}{1.000pt}}
\multiput(782.00,644.92)(-2.000,-2.000){2}{\rule{0.482pt}{1.000pt}}
\put(776,641.92){\rule{0.964pt}{1.000pt}}
\multiput(778.00,642.92)(-2.000,-2.000){2}{\rule{0.482pt}{1.000pt}}
\put(772,639.92){\rule{0.964pt}{1.000pt}}
\multiput(774.00,640.92)(-2.000,-2.000){2}{\rule{0.482pt}{1.000pt}}
\put(767,637.92){\rule{1.204pt}{1.000pt}}
\multiput(769.50,638.92)(-2.500,-2.000){2}{\rule{0.602pt}{1.000pt}}
\put(763,635.92){\rule{0.964pt}{1.000pt}}
\multiput(765.00,636.92)(-2.000,-2.000){2}{\rule{0.482pt}{1.000pt}}
\put(759,634.42){\rule{0.964pt}{1.000pt}}
\multiput(761.00,634.92)(-2.000,-1.000){2}{\rule{0.482pt}{1.000pt}}
\put(755,632.92){\rule{0.964pt}{1.000pt}}
\multiput(757.00,633.92)(-2.000,-2.000){2}{\rule{0.482pt}{1.000pt}}
\put(750,631.42){\rule{1.204pt}{1.000pt}}
\multiput(752.50,631.92)(-2.500,-1.000){2}{\rule{0.602pt}{1.000pt}}
\put(746,629.92){\rule{0.964pt}{1.000pt}}
\multiput(748.00,630.92)(-2.000,-2.000){2}{\rule{0.482pt}{1.000pt}}
\put(741,628.42){\rule{1.204pt}{1.000pt}}
\multiput(743.50,628.92)(-2.500,-1.000){2}{\rule{0.602pt}{1.000pt}}
\put(737,627.42){\rule{0.964pt}{1.000pt}}
\multiput(739.00,627.92)(-2.000,-1.000){2}{\rule{0.482pt}{1.000pt}}
\put(732,625.92){\rule{1.204pt}{1.000pt}}
\multiput(734.50,626.92)(-2.500,-2.000){2}{\rule{0.602pt}{1.000pt}}
\put(728,624.42){\rule{0.964pt}{1.000pt}}
\multiput(730.00,624.92)(-2.000,-1.000){2}{\rule{0.482pt}{1.000pt}}
\put(723,623.42){\rule{1.204pt}{1.000pt}}
\multiput(725.50,623.92)(-2.500,-1.000){2}{\rule{0.602pt}{1.000pt}}
\put(786.0,637.0){\rule[-0.500pt]{1.000pt}{2.168pt}}
\put(714,622.42){\rule{1.204pt}{1.000pt}}
\multiput(716.50,622.92)(-2.500,-1.000){2}{\rule{0.602pt}{1.000pt}}
\put(710,621.42){\rule{0.964pt}{1.000pt}}
\multiput(712.00,621.92)(-2.000,-1.000){2}{\rule{0.482pt}{1.000pt}}
\put(705,620.42){\rule{1.204pt}{1.000pt}}
\multiput(707.50,620.92)(-2.500,-1.000){2}{\rule{0.602pt}{1.000pt}}
\put(719.0,625.0){\usebox{\plotpoint}}
\put(696,619.42){\rule{1.204pt}{1.000pt}}
\multiput(698.50,619.92)(-2.500,-1.000){2}{\rule{0.602pt}{1.000pt}}
\put(701.0,622.0){\usebox{\plotpoint}}
\put(650,619.42){\rule{0.964pt}{1.000pt}}
\multiput(652.00,618.92)(-2.000,1.000){2}{\rule{0.482pt}{1.000pt}}
\put(654.0,621.0){\rule[-0.500pt]{10.118pt}{1.000pt}}
\put(640,620.42){\rule{1.204pt}{1.000pt}}
\multiput(642.50,619.92)(-2.500,1.000){2}{\rule{0.602pt}{1.000pt}}
\put(636,621.42){\rule{0.964pt}{1.000pt}}
\multiput(638.00,620.92)(-2.000,1.000){2}{\rule{0.482pt}{1.000pt}}
\put(631,622.42){\rule{1.204pt}{1.000pt}}
\multiput(633.50,621.92)(-2.500,1.000){2}{\rule{0.602pt}{1.000pt}}
\put(645.0,622.0){\rule[-0.500pt]{1.204pt}{1.000pt}}
\put(622,623.42){\rule{1.204pt}{1.000pt}}
\multiput(624.50,622.92)(-2.500,1.000){2}{\rule{0.602pt}{1.000pt}}
\put(618,624.42){\rule{0.964pt}{1.000pt}}
\multiput(620.00,623.92)(-2.000,1.000){2}{\rule{0.482pt}{1.000pt}}
\put(613,625.92){\rule{1.204pt}{1.000pt}}
\multiput(615.50,624.92)(-2.500,2.000){2}{\rule{0.602pt}{1.000pt}}
\put(609,627.42){\rule{0.964pt}{1.000pt}}
\multiput(611.00,626.92)(-2.000,1.000){2}{\rule{0.482pt}{1.000pt}}
\put(604,628.42){\rule{1.204pt}{1.000pt}}
\multiput(606.50,627.92)(-2.500,1.000){2}{\rule{0.602pt}{1.000pt}}
\put(600,629.92){\rule{0.964pt}{1.000pt}}
\multiput(602.00,628.92)(-2.000,2.000){2}{\rule{0.482pt}{1.000pt}}
\put(596,631.42){\rule{0.964pt}{1.000pt}}
\multiput(598.00,630.92)(-2.000,1.000){2}{\rule{0.482pt}{1.000pt}}
\put(591,632.92){\rule{1.204pt}{1.000pt}}
\multiput(593.50,631.92)(-2.500,2.000){2}{\rule{0.602pt}{1.000pt}}
\put(587,634.42){\rule{0.964pt}{1.000pt}}
\multiput(589.00,633.92)(-2.000,1.000){2}{\rule{0.482pt}{1.000pt}}
\put(583,635.92){\rule{0.964pt}{1.000pt}}
\multiput(585.00,634.92)(-2.000,2.000){2}{\rule{0.482pt}{1.000pt}}
\put(579,637.92){\rule{0.964pt}{1.000pt}}
\multiput(581.00,636.92)(-2.000,2.000){2}{\rule{0.482pt}{1.000pt}}
\put(574,639.92){\rule{1.204pt}{1.000pt}}
\multiput(576.50,638.92)(-2.500,2.000){2}{\rule{0.602pt}{1.000pt}}
\put(570,641.92){\rule{0.964pt}{1.000pt}}
\multiput(572.00,640.92)(-2.000,2.000){2}{\rule{0.482pt}{1.000pt}}
\put(566,643.92){\rule{0.964pt}{1.000pt}}
\multiput(568.00,642.92)(-2.000,2.000){2}{\rule{0.482pt}{1.000pt}}
\put(564,644.42){\rule{0.482pt}{1.000pt}}
\multiput(565.00,644.92)(-1.000,-1.000){2}{\rule{0.241pt}{1.000pt}}
\put(627.0,625.0){\usebox{\plotpoint}}
\put(561.42,629){\rule{1.000pt}{0.964pt}}
\multiput(561.92,631.00)(-1.000,-2.000){2}{\rule{1.000pt}{0.482pt}}
\put(564.0,633.0){\rule[-0.500pt]{1.000pt}{3.132pt}}
\put(560.42,620){\rule{1.000pt}{0.964pt}}
\multiput(560.92,622.00)(-1.000,-2.000){2}{\rule{1.000pt}{0.482pt}}
\put(559.42,616){\rule{1.000pt}{0.964pt}}
\multiput(559.92,618.00)(-1.000,-2.000){2}{\rule{1.000pt}{0.482pt}}
\put(563.0,624.0){\rule[-0.500pt]{1.000pt}{1.204pt}}
\put(558.42,608){\rule{1.000pt}{0.964pt}}
\multiput(558.92,610.00)(-1.000,-2.000){2}{\rule{1.000pt}{0.482pt}}
\put(557.42,604){\rule{1.000pt}{0.964pt}}
\multiput(557.92,606.00)(-1.000,-2.000){2}{\rule{1.000pt}{0.482pt}}
\put(556.42,599){\rule{1.000pt}{1.204pt}}
\multiput(556.92,601.50)(-1.000,-2.500){2}{\rule{1.000pt}{0.602pt}}
\put(555.42,595){\rule{1.000pt}{0.964pt}}
\multiput(555.92,597.00)(-1.000,-2.000){2}{\rule{1.000pt}{0.482pt}}
\put(553.92,591){\rule{1.000pt}{0.964pt}}
\multiput(554.92,593.00)(-2.000,-2.000){2}{\rule{1.000pt}{0.482pt}}
\put(552.42,587){\rule{1.000pt}{0.964pt}}
\multiput(552.92,589.00)(-1.000,-2.000){2}{\rule{1.000pt}{0.482pt}}
\put(551.42,583){\rule{1.000pt}{0.964pt}}
\multiput(551.92,585.00)(-1.000,-2.000){2}{\rule{1.000pt}{0.482pt}}
\put(549.92,579){\rule{1.000pt}{0.964pt}}
\multiput(550.92,581.00)(-2.000,-2.000){2}{\rule{1.000pt}{0.482pt}}
\put(547.92,575){\rule{1.000pt}{0.964pt}}
\multiput(548.92,577.00)(-2.000,-2.000){2}{\rule{1.000pt}{0.482pt}}
\put(546.42,571){\rule{1.000pt}{0.964pt}}
\multiput(546.92,573.00)(-1.000,-2.000){2}{\rule{1.000pt}{0.482pt}}
\put(544.92,567){\rule{1.000pt}{0.964pt}}
\multiput(545.92,569.00)(-2.000,-2.000){2}{\rule{1.000pt}{0.482pt}}
\put(542.92,563){\rule{1.000pt}{0.964pt}}
\multiput(543.92,565.00)(-2.000,-2.000){2}{\rule{1.000pt}{0.482pt}}
\put(540.92,560){\rule{1.000pt}{0.723pt}}
\multiput(541.92,561.50)(-2.000,-1.500){2}{\rule{1.000pt}{0.361pt}}
\put(538.92,556){\rule{1.000pt}{0.964pt}}
\multiput(539.92,558.00)(-2.000,-2.000){2}{\rule{1.000pt}{0.482pt}}
\put(536.92,552){\rule{1.000pt}{0.964pt}}
\multiput(537.92,554.00)(-2.000,-2.000){2}{\rule{1.000pt}{0.482pt}}
\put(534.92,548){\rule{1.000pt}{0.964pt}}
\multiput(535.92,550.00)(-2.000,-2.000){2}{\rule{1.000pt}{0.482pt}}
\put(533,544.42){\rule{0.723pt}{1.000pt}}
\multiput(534.50,545.92)(-1.500,-3.000){2}{\rule{0.361pt}{1.000pt}}
\put(529.92,541){\rule{1.000pt}{0.964pt}}
\multiput(530.92,543.00)(-2.000,-2.000){2}{\rule{1.000pt}{0.482pt}}
\put(528,537.42){\rule{0.723pt}{1.000pt}}
\multiput(529.50,538.92)(-1.500,-3.000){2}{\rule{0.361pt}{1.000pt}}
\put(524.92,534){\rule{1.000pt}{0.964pt}}
\multiput(525.92,536.00)(-2.000,-2.000){2}{\rule{1.000pt}{0.482pt}}
\put(522.42,530){\rule{1.000pt}{0.964pt}}
\multiput(523.92,532.00)(-3.000,-2.000){2}{\rule{1.000pt}{0.482pt}}
\put(520,526.42){\rule{0.723pt}{1.000pt}}
\multiput(521.50,527.92)(-1.500,-3.000){2}{\rule{0.361pt}{1.000pt}}
\put(516.92,524){\rule{1.000pt}{0.723pt}}
\multiput(517.92,525.50)(-2.000,-1.500){2}{\rule{1.000pt}{0.361pt}}
\put(514.42,520){\rule{1.000pt}{0.964pt}}
\multiput(515.92,522.00)(-3.000,-2.000){2}{\rule{1.000pt}{0.482pt}}
\put(512,516.42){\rule{0.723pt}{1.000pt}}
\multiput(513.50,517.92)(-1.500,-3.000){2}{\rule{0.361pt}{1.000pt}}
\put(509,513.42){\rule{0.723pt}{1.000pt}}
\multiput(510.50,514.92)(-1.500,-3.000){2}{\rule{0.361pt}{1.000pt}}
\put(506,510.42){\rule{0.723pt}{1.000pt}}
\multiput(507.50,511.92)(-1.500,-3.000){2}{\rule{0.361pt}{1.000pt}}
\put(502,507.42){\rule{0.964pt}{1.000pt}}
\multiput(504.00,508.92)(-2.000,-3.000){2}{\rule{0.482pt}{1.000pt}}
\put(499,504.42){\rule{0.723pt}{1.000pt}}
\multiput(500.50,505.92)(-1.500,-3.000){2}{\rule{0.361pt}{1.000pt}}
\put(496,501.42){\rule{0.723pt}{1.000pt}}
\multiput(497.50,502.92)(-1.500,-3.000){2}{\rule{0.361pt}{1.000pt}}
\put(492,498.42){\rule{0.964pt}{1.000pt}}
\multiput(494.00,499.92)(-2.000,-3.000){2}{\rule{0.482pt}{1.000pt}}
\put(489,495.42){\rule{0.723pt}{1.000pt}}
\multiput(490.50,496.92)(-1.500,-3.000){2}{\rule{0.361pt}{1.000pt}}
\put(486,492.42){\rule{0.723pt}{1.000pt}}
\multiput(487.50,493.92)(-1.500,-3.000){2}{\rule{0.361pt}{1.000pt}}
\put(482,489.42){\rule{0.964pt}{1.000pt}}
\multiput(484.00,490.92)(-2.000,-3.000){2}{\rule{0.482pt}{1.000pt}}
\put(478,486.92){\rule{0.964pt}{1.000pt}}
\multiput(480.00,487.92)(-2.000,-2.000){2}{\rule{0.482pt}{1.000pt}}
\put(475,484.42){\rule{0.723pt}{1.000pt}}
\multiput(476.50,485.92)(-1.500,-3.000){2}{\rule{0.361pt}{1.000pt}}
\put(471,481.92){\rule{0.964pt}{1.000pt}}
\multiput(473.00,482.92)(-2.000,-2.000){2}{\rule{0.482pt}{1.000pt}}
\put(467,479.42){\rule{0.964pt}{1.000pt}}
\multiput(469.00,480.92)(-2.000,-3.000){2}{\rule{0.482pt}{1.000pt}}
\put(463,476.92){\rule{0.964pt}{1.000pt}}
\multiput(465.00,477.92)(-2.000,-2.000){2}{\rule{0.482pt}{1.000pt}}
\put(459,474.92){\rule{0.964pt}{1.000pt}}
\multiput(461.00,475.92)(-2.000,-2.000){2}{\rule{0.482pt}{1.000pt}}
\put(455,472.92){\rule{0.964pt}{1.000pt}}
\multiput(457.00,473.92)(-2.000,-2.000){2}{\rule{0.482pt}{1.000pt}}
\put(561.0,612.0){\usebox{\plotpoint}}
\put(455,467.92){\rule{0.964pt}{1.000pt}}
\multiput(455.00,468.92)(2.000,-2.000){2}{\rule{0.482pt}{1.000pt}}
\put(459,465.92){\rule{0.964pt}{1.000pt}}
\multiput(459.00,466.92)(2.000,-2.000){2}{\rule{0.482pt}{1.000pt}}
\put(463,463.92){\rule{0.964pt}{1.000pt}}
\multiput(463.00,464.92)(2.000,-2.000){2}{\rule{0.482pt}{1.000pt}}
\put(467,461.42){\rule{0.964pt}{1.000pt}}
\multiput(467.00,462.92)(2.000,-3.000){2}{\rule{0.482pt}{1.000pt}}
\put(471,458.92){\rule{0.964pt}{1.000pt}}
\multiput(471.00,459.92)(2.000,-2.000){2}{\rule{0.482pt}{1.000pt}}
\put(475,456.42){\rule{0.723pt}{1.000pt}}
\multiput(475.00,457.92)(1.500,-3.000){2}{\rule{0.361pt}{1.000pt}}
\put(478,453.92){\rule{0.964pt}{1.000pt}}
\multiput(478.00,454.92)(2.000,-2.000){2}{\rule{0.482pt}{1.000pt}}
\put(482,451.42){\rule{0.964pt}{1.000pt}}
\multiput(482.00,452.92)(2.000,-3.000){2}{\rule{0.482pt}{1.000pt}}
\put(486,448.42){\rule{0.723pt}{1.000pt}}
\multiput(486.00,449.92)(1.500,-3.000){2}{\rule{0.361pt}{1.000pt}}
\put(489,445.42){\rule{0.723pt}{1.000pt}}
\multiput(489.00,446.92)(1.500,-3.000){2}{\rule{0.361pt}{1.000pt}}
\put(492,442.42){\rule{0.964pt}{1.000pt}}
\multiput(492.00,443.92)(2.000,-3.000){2}{\rule{0.482pt}{1.000pt}}
\put(496,439.42){\rule{0.723pt}{1.000pt}}
\multiput(496.00,440.92)(1.500,-3.000){2}{\rule{0.361pt}{1.000pt}}
\put(499,436.42){\rule{0.723pt}{1.000pt}}
\multiput(499.00,437.92)(1.500,-3.000){2}{\rule{0.361pt}{1.000pt}}
\put(502,433.42){\rule{0.964pt}{1.000pt}}
\multiput(502.00,434.92)(2.000,-3.000){2}{\rule{0.482pt}{1.000pt}}
\put(506,430.42){\rule{0.723pt}{1.000pt}}
\multiput(506.00,431.92)(1.500,-3.000){2}{\rule{0.361pt}{1.000pt}}
\put(509,427.42){\rule{0.723pt}{1.000pt}}
\multiput(509.00,428.92)(1.500,-3.000){2}{\rule{0.361pt}{1.000pt}}
\put(512,424.42){\rule{0.723pt}{1.000pt}}
\multiput(512.00,425.92)(1.500,-3.000){2}{\rule{0.361pt}{1.000pt}}
\put(514.42,421){\rule{1.000pt}{0.964pt}}
\multiput(512.92,423.00)(3.000,-2.000){2}{\rule{1.000pt}{0.482pt}}
\put(516.92,418){\rule{1.000pt}{0.723pt}}
\multiput(515.92,419.50)(2.000,-1.500){2}{\rule{1.000pt}{0.361pt}}
\put(520,414.42){\rule{0.723pt}{1.000pt}}
\multiput(520.00,415.92)(1.500,-3.000){2}{\rule{0.361pt}{1.000pt}}
\put(522.42,411){\rule{1.000pt}{0.964pt}}
\multiput(520.92,413.00)(3.000,-2.000){2}{\rule{1.000pt}{0.482pt}}
\put(524.92,407){\rule{1.000pt}{0.964pt}}
\multiput(523.92,409.00)(2.000,-2.000){2}{\rule{1.000pt}{0.482pt}}
\put(528,403.42){\rule{0.723pt}{1.000pt}}
\multiput(528.00,404.92)(1.500,-3.000){2}{\rule{0.361pt}{1.000pt}}
\put(529.92,400){\rule{1.000pt}{0.964pt}}
\multiput(528.92,402.00)(2.000,-2.000){2}{\rule{1.000pt}{0.482pt}}
\put(533,396.42){\rule{0.723pt}{1.000pt}}
\multiput(533.00,397.92)(1.500,-3.000){2}{\rule{0.361pt}{1.000pt}}
\put(534.92,393){\rule{1.000pt}{0.964pt}}
\multiput(533.92,395.00)(2.000,-2.000){2}{\rule{1.000pt}{0.482pt}}
\put(536.92,389){\rule{1.000pt}{0.964pt}}
\multiput(535.92,391.00)(2.000,-2.000){2}{\rule{1.000pt}{0.482pt}}
\put(538.92,385){\rule{1.000pt}{0.964pt}}
\multiput(537.92,387.00)(2.000,-2.000){2}{\rule{1.000pt}{0.482pt}}
\put(540.92,382){\rule{1.000pt}{0.723pt}}
\multiput(539.92,383.50)(2.000,-1.500){2}{\rule{1.000pt}{0.361pt}}
\put(542.92,378){\rule{1.000pt}{0.964pt}}
\multiput(541.92,380.00)(2.000,-2.000){2}{\rule{1.000pt}{0.482pt}}
\put(544.92,374){\rule{1.000pt}{0.964pt}}
\multiput(543.92,376.00)(2.000,-2.000){2}{\rule{1.000pt}{0.482pt}}
\put(546.42,370){\rule{1.000pt}{0.964pt}}
\multiput(545.92,372.00)(1.000,-2.000){2}{\rule{1.000pt}{0.482pt}}
\put(547.92,366){\rule{1.000pt}{0.964pt}}
\multiput(546.92,368.00)(2.000,-2.000){2}{\rule{1.000pt}{0.482pt}}
\put(549.92,362){\rule{1.000pt}{0.964pt}}
\multiput(548.92,364.00)(2.000,-2.000){2}{\rule{1.000pt}{0.482pt}}
\put(551.42,358){\rule{1.000pt}{0.964pt}}
\multiput(550.92,360.00)(1.000,-2.000){2}{\rule{1.000pt}{0.482pt}}
\put(552.42,354){\rule{1.000pt}{0.964pt}}
\multiput(551.92,356.00)(1.000,-2.000){2}{\rule{1.000pt}{0.482pt}}
\put(553.92,350){\rule{1.000pt}{0.964pt}}
\multiput(552.92,352.00)(2.000,-2.000){2}{\rule{1.000pt}{0.482pt}}
\put(555.42,346){\rule{1.000pt}{0.964pt}}
\multiput(554.92,348.00)(1.000,-2.000){2}{\rule{1.000pt}{0.482pt}}
\put(556.42,341){\rule{1.000pt}{1.204pt}}
\multiput(555.92,343.50)(1.000,-2.500){2}{\rule{1.000pt}{0.602pt}}
\put(557.42,337){\rule{1.000pt}{0.964pt}}
\multiput(556.92,339.00)(1.000,-2.000){2}{\rule{1.000pt}{0.482pt}}
\put(558.42,333){\rule{1.000pt}{0.964pt}}
\multiput(557.92,335.00)(1.000,-2.000){2}{\rule{1.000pt}{0.482pt}}
\put(455.0,471.0){\usebox{\plotpoint}}
\put(559.42,325){\rule{1.000pt}{0.964pt}}
\multiput(558.92,327.00)(1.000,-2.000){2}{\rule{1.000pt}{0.482pt}}
\put(560.42,321){\rule{1.000pt}{0.964pt}}
\multiput(559.92,323.00)(1.000,-2.000){2}{\rule{1.000pt}{0.482pt}}
\put(561.0,329.0){\usebox{\plotpoint}}
\put(561.42,312){\rule{1.000pt}{0.964pt}}
\multiput(560.92,314.00)(1.000,-2.000){2}{\rule{1.000pt}{0.482pt}}
\put(563.0,316.0){\rule[-0.500pt]{1.000pt}{1.204pt}}
\put(564,296.42){\rule{0.482pt}{1.000pt}}
\multiput(564.00,296.92)(1.000,-1.000){2}{\rule{0.241pt}{1.000pt}}
\put(566,296.92){\rule{0.964pt}{1.000pt}}
\multiput(566.00,295.92)(2.000,2.000){2}{\rule{0.482pt}{1.000pt}}
\put(570,298.92){\rule{0.964pt}{1.000pt}}
\multiput(570.00,297.92)(2.000,2.000){2}{\rule{0.482pt}{1.000pt}}
\put(574,300.92){\rule{1.204pt}{1.000pt}}
\multiput(574.00,299.92)(2.500,2.000){2}{\rule{0.602pt}{1.000pt}}
\put(579,302.92){\rule{0.964pt}{1.000pt}}
\multiput(579.00,301.92)(2.000,2.000){2}{\rule{0.482pt}{1.000pt}}
\put(583,304.92){\rule{0.964pt}{1.000pt}}
\multiput(583.00,303.92)(2.000,2.000){2}{\rule{0.482pt}{1.000pt}}
\put(587,306.42){\rule{0.964pt}{1.000pt}}
\multiput(587.00,305.92)(2.000,1.000){2}{\rule{0.482pt}{1.000pt}}
\put(591,307.92){\rule{1.204pt}{1.000pt}}
\multiput(591.00,306.92)(2.500,2.000){2}{\rule{0.602pt}{1.000pt}}
\put(596,309.42){\rule{0.964pt}{1.000pt}}
\multiput(596.00,308.92)(2.000,1.000){2}{\rule{0.482pt}{1.000pt}}
\put(600,310.92){\rule{0.964pt}{1.000pt}}
\multiput(600.00,309.92)(2.000,2.000){2}{\rule{0.482pt}{1.000pt}}
\put(604,312.42){\rule{1.204pt}{1.000pt}}
\multiput(604.00,311.92)(2.500,1.000){2}{\rule{0.602pt}{1.000pt}}
\put(609,313.42){\rule{0.964pt}{1.000pt}}
\multiput(609.00,312.92)(2.000,1.000){2}{\rule{0.482pt}{1.000pt}}
\put(613,314.92){\rule{1.204pt}{1.000pt}}
\multiput(613.00,313.92)(2.500,2.000){2}{\rule{0.602pt}{1.000pt}}
\put(618,316.42){\rule{0.964pt}{1.000pt}}
\multiput(618.00,315.92)(2.000,1.000){2}{\rule{0.482pt}{1.000pt}}
\put(622,317.42){\rule{1.204pt}{1.000pt}}
\multiput(622.00,316.92)(2.500,1.000){2}{\rule{0.602pt}{1.000pt}}
\put(564.0,299.0){\rule[-0.500pt]{1.000pt}{3.132pt}}
\put(631,318.42){\rule{1.204pt}{1.000pt}}
\multiput(631.00,317.92)(2.500,1.000){2}{\rule{0.602pt}{1.000pt}}
\put(636,319.42){\rule{0.964pt}{1.000pt}}
\multiput(636.00,318.92)(2.000,1.000){2}{\rule{0.482pt}{1.000pt}}
\put(640,320.42){\rule{1.204pt}{1.000pt}}
\multiput(640.00,319.92)(2.500,1.000){2}{\rule{0.602pt}{1.000pt}}
\put(627.0,320.0){\usebox{\plotpoint}}
\put(650,321.42){\rule{0.964pt}{1.000pt}}
\multiput(650.00,320.92)(2.000,1.000){2}{\rule{0.482pt}{1.000pt}}
\put(645.0,323.0){\rule[-0.500pt]{1.204pt}{1.000pt}}
\put(696,321.42){\rule{1.204pt}{1.000pt}}
\multiput(696.00,321.92)(2.500,-1.000){2}{\rule{0.602pt}{1.000pt}}
\put(654.0,324.0){\rule[-0.500pt]{10.118pt}{1.000pt}}
\put(705,320.42){\rule{1.204pt}{1.000pt}}
\multiput(705.00,320.92)(2.500,-1.000){2}{\rule{0.602pt}{1.000pt}}
\put(710,319.42){\rule{0.964pt}{1.000pt}}
\multiput(710.00,319.92)(2.000,-1.000){2}{\rule{0.482pt}{1.000pt}}
\put(714,318.42){\rule{1.204pt}{1.000pt}}
\multiput(714.00,318.92)(2.500,-1.000){2}{\rule{0.602pt}{1.000pt}}
\put(701.0,323.0){\usebox{\plotpoint}}
\put(723,317.42){\rule{1.204pt}{1.000pt}}
\multiput(723.00,317.92)(2.500,-1.000){2}{\rule{0.602pt}{1.000pt}}
\put(728,316.42){\rule{0.964pt}{1.000pt}}
\multiput(728.00,316.92)(2.000,-1.000){2}{\rule{0.482pt}{1.000pt}}
\put(732,314.92){\rule{1.204pt}{1.000pt}}
\multiput(732.00,315.92)(2.500,-2.000){2}{\rule{0.602pt}{1.000pt}}
\put(737,313.42){\rule{0.964pt}{1.000pt}}
\multiput(737.00,313.92)(2.000,-1.000){2}{\rule{0.482pt}{1.000pt}}
\put(741,312.42){\rule{1.204pt}{1.000pt}}
\multiput(741.00,312.92)(2.500,-1.000){2}{\rule{0.602pt}{1.000pt}}
\put(746,310.92){\rule{0.964pt}{1.000pt}}
\multiput(746.00,311.92)(2.000,-2.000){2}{\rule{0.482pt}{1.000pt}}
\put(750,309.42){\rule{1.204pt}{1.000pt}}
\multiput(750.00,309.92)(2.500,-1.000){2}{\rule{0.602pt}{1.000pt}}
\put(755,307.92){\rule{0.964pt}{1.000pt}}
\multiput(755.00,308.92)(2.000,-2.000){2}{\rule{0.482pt}{1.000pt}}
\put(759,306.42){\rule{0.964pt}{1.000pt}}
\multiput(759.00,306.92)(2.000,-1.000){2}{\rule{0.482pt}{1.000pt}}
\put(763,304.92){\rule{0.964pt}{1.000pt}}
\multiput(763.00,305.92)(2.000,-2.000){2}{\rule{0.482pt}{1.000pt}}
\put(767,302.92){\rule{1.204pt}{1.000pt}}
\multiput(767.00,303.92)(2.500,-2.000){2}{\rule{0.602pt}{1.000pt}}
\put(772,300.92){\rule{0.964pt}{1.000pt}}
\multiput(772.00,301.92)(2.000,-2.000){2}{\rule{0.482pt}{1.000pt}}
\put(776,298.92){\rule{0.964pt}{1.000pt}}
\multiput(776.00,299.92)(2.000,-2.000){2}{\rule{0.482pt}{1.000pt}}
\put(780,296.92){\rule{0.964pt}{1.000pt}}
\multiput(780.00,297.92)(2.000,-2.000){2}{\rule{0.482pt}{1.000pt}}
\put(784,296.42){\rule{0.482pt}{1.000pt}}
\multiput(784.00,295.92)(1.000,1.000){2}{\rule{0.241pt}{1.000pt}}
\put(719.0,320.0){\usebox{\plotpoint}}
\put(784.42,308){\rule{1.000pt}{0.964pt}}
\multiput(783.92,308.00)(1.000,2.000){2}{\rule{1.000pt}{0.482pt}}
\put(786.0,299.0){\rule[-0.500pt]{1.000pt}{2.168pt}}
\put(785.42,321){\rule{1.000pt}{0.964pt}}
\multiput(784.92,321.00)(1.000,2.000){2}{\rule{1.000pt}{0.482pt}}
\put(786.42,325){\rule{1.000pt}{0.964pt}}
\multiput(785.92,325.00)(1.000,2.000){2}{\rule{1.000pt}{0.482pt}}
\put(787.0,312.0){\rule[-0.500pt]{1.000pt}{2.168pt}}
\put(787.42,333){\rule{1.000pt}{0.964pt}}
\multiput(786.92,333.00)(1.000,2.000){2}{\rule{1.000pt}{0.482pt}}
\put(788.42,337){\rule{1.000pt}{0.964pt}}
\multiput(787.92,337.00)(1.000,2.000){2}{\rule{1.000pt}{0.482pt}}
\put(789.42,341){\rule{1.000pt}{1.204pt}}
\multiput(788.92,341.00)(1.000,2.500){2}{\rule{1.000pt}{0.602pt}}
\put(790.92,346){\rule{1.000pt}{0.964pt}}
\multiput(789.92,346.00)(2.000,2.000){2}{\rule{1.000pt}{0.482pt}}
\put(792.42,350){\rule{1.000pt}{0.964pt}}
\multiput(791.92,350.00)(1.000,2.000){2}{\rule{1.000pt}{0.482pt}}
\put(793.42,354){\rule{1.000pt}{0.964pt}}
\multiput(792.92,354.00)(1.000,2.000){2}{\rule{1.000pt}{0.482pt}}
\put(794.92,358){\rule{1.000pt}{0.964pt}}
\multiput(793.92,358.00)(2.000,2.000){2}{\rule{1.000pt}{0.482pt}}
\put(796.42,362){\rule{1.000pt}{0.964pt}}
\multiput(795.92,362.00)(1.000,2.000){2}{\rule{1.000pt}{0.482pt}}
\put(797.92,366){\rule{1.000pt}{0.964pt}}
\multiput(796.92,366.00)(2.000,2.000){2}{\rule{1.000pt}{0.482pt}}
\put(799.42,370){\rule{1.000pt}{0.964pt}}
\multiput(798.92,370.00)(1.000,2.000){2}{\rule{1.000pt}{0.482pt}}
\put(800.92,374){\rule{1.000pt}{0.964pt}}
\multiput(799.92,374.00)(2.000,2.000){2}{\rule{1.000pt}{0.482pt}}
\put(802.92,378){\rule{1.000pt}{0.964pt}}
\multiput(801.92,378.00)(2.000,2.000){2}{\rule{1.000pt}{0.482pt}}
\put(804.92,382){\rule{1.000pt}{0.723pt}}
\multiput(803.92,382.00)(2.000,1.500){2}{\rule{1.000pt}{0.361pt}}
\put(806.92,385){\rule{1.000pt}{0.964pt}}
\multiput(805.92,385.00)(2.000,2.000){2}{\rule{1.000pt}{0.482pt}}
\put(808.92,389){\rule{1.000pt}{0.964pt}}
\multiput(807.92,389.00)(2.000,2.000){2}{\rule{1.000pt}{0.482pt}}
\put(811.42,393){\rule{1.000pt}{0.964pt}}
\multiput(809.92,393.00)(3.000,2.000){2}{\rule{1.000pt}{0.482pt}}
\put(813.92,397){\rule{1.000pt}{0.723pt}}
\multiput(812.92,397.00)(2.000,1.500){2}{\rule{1.000pt}{0.361pt}}
\put(815.92,400){\rule{1.000pt}{0.964pt}}
\multiput(814.92,400.00)(2.000,2.000){2}{\rule{1.000pt}{0.482pt}}
\put(819,403.42){\rule{0.723pt}{1.000pt}}
\multiput(819.00,401.92)(1.500,3.000){2}{\rule{0.361pt}{1.000pt}}
\put(820.92,407){\rule{1.000pt}{0.964pt}}
\multiput(819.92,407.00)(2.000,2.000){2}{\rule{1.000pt}{0.482pt}}
\put(823.42,411){\rule{1.000pt}{0.964pt}}
\multiput(821.92,411.00)(3.000,2.000){2}{\rule{1.000pt}{0.482pt}}
\put(827,414.42){\rule{0.723pt}{1.000pt}}
\multiput(827.00,412.92)(1.500,3.000){2}{\rule{0.361pt}{1.000pt}}
\put(828.92,418){\rule{1.000pt}{0.723pt}}
\multiput(827.92,418.00)(2.000,1.500){2}{\rule{1.000pt}{0.361pt}}
\put(831.42,421){\rule{1.000pt}{0.964pt}}
\multiput(829.92,421.00)(3.000,2.000){2}{\rule{1.000pt}{0.482pt}}
\put(835,424.42){\rule{0.723pt}{1.000pt}}
\multiput(835.00,422.92)(1.500,3.000){2}{\rule{0.361pt}{1.000pt}}
\put(838,427.42){\rule{0.723pt}{1.000pt}}
\multiput(838.00,425.92)(1.500,3.000){2}{\rule{0.361pt}{1.000pt}}
\put(841,430.42){\rule{0.723pt}{1.000pt}}
\multiput(841.00,428.92)(1.500,3.000){2}{\rule{0.361pt}{1.000pt}}
\put(844,433.42){\rule{0.964pt}{1.000pt}}
\multiput(844.00,431.92)(2.000,3.000){2}{\rule{0.482pt}{1.000pt}}
\put(848,436.42){\rule{0.723pt}{1.000pt}}
\multiput(848.00,434.92)(1.500,3.000){2}{\rule{0.361pt}{1.000pt}}
\put(851,439.42){\rule{0.723pt}{1.000pt}}
\multiput(851.00,437.92)(1.500,3.000){2}{\rule{0.361pt}{1.000pt}}
\put(854,442.42){\rule{0.964pt}{1.000pt}}
\multiput(854.00,440.92)(2.000,3.000){2}{\rule{0.482pt}{1.000pt}}
\put(858,445.42){\rule{0.723pt}{1.000pt}}
\multiput(858.00,443.92)(1.500,3.000){2}{\rule{0.361pt}{1.000pt}}
\put(861,448.42){\rule{0.964pt}{1.000pt}}
\multiput(861.00,446.92)(2.000,3.000){2}{\rule{0.482pt}{1.000pt}}
\put(865,451.42){\rule{0.723pt}{1.000pt}}
\multiput(865.00,449.92)(1.500,3.000){2}{\rule{0.361pt}{1.000pt}}
\put(868,453.92){\rule{0.964pt}{1.000pt}}
\multiput(868.00,452.92)(2.000,2.000){2}{\rule{0.482pt}{1.000pt}}
\put(872,456.42){\rule{0.964pt}{1.000pt}}
\multiput(872.00,454.92)(2.000,3.000){2}{\rule{0.482pt}{1.000pt}}
\put(876,458.92){\rule{0.723pt}{1.000pt}}
\multiput(876.00,457.92)(1.500,2.000){2}{\rule{0.361pt}{1.000pt}}
\put(879,461.42){\rule{0.964pt}{1.000pt}}
\multiput(879.00,459.92)(2.000,3.000){2}{\rule{0.482pt}{1.000pt}}
\put(883,463.92){\rule{0.964pt}{1.000pt}}
\multiput(883.00,462.92)(2.000,2.000){2}{\rule{0.482pt}{1.000pt}}
\put(887,465.92){\rule{0.964pt}{1.000pt}}
\multiput(887.00,464.92)(2.000,2.000){2}{\rule{0.482pt}{1.000pt}}
\put(891,467.92){\rule{0.964pt}{1.000pt}}
\multiput(891.00,466.92)(2.000,2.000){2}{\rule{0.482pt}{1.000pt}}
\put(789.0,329.0){\usebox{\plotpoint}}
\end{picture}

The disappointing conclusion above is based on the assumption
that the model manifold for a string trajectory describing the
elementary process of joining or splitting is given by a hexagon
with smooth "timelike" boundaries. If we consider such a process
in the Minkowski target space the boundary of the
corresponding word sheet
has a corner at the interaction point. It follows that if we
choose the hyperbolic hexagon (Fig.1.a) as a model manifold the causal
string trajectories will be described by singular (at the interaction
point) functions. An equivalent description in terms of
regular $x$-functions can be achieved if we choose the
"light-cone" model manifold (Fig.1.b). Note that in the
"light-cone" model manifold the internal angle of the
"interaction" corner equals $2\pi$. This is related to the
assumption that no particular interaction occurs at the joining
point. Applying the reasoning based on the Gauss-Bonnet theorem
in the case of "light-cone" model manifold one gets as in the
free theory the vanishing cosmological constant.

It should be stressed that as far as the critical string theory
is concerned the choice of the model manifold is irrelevant.
In fact due to the decoupling of the conformal factor the
critical on-shell amplitude is invariant with respect to the
conformal transformations of the model manifold.
As it was mentioned in the introduction this is one of the  features
crucial for the equivalence of the Polyakov dual model
with the relativistic string theory in the critical dimension.

The situation in the noncritical string theory is
different. Roughly speaking due to the fact that the
boundary conditions for the conformal factor are fixed
the Liouville action "hears" the
shape of the model manifold and the theories based on the
hyperbolic hexagon and on the "light-cone" model manifold are
different.

It follows from the considerations above that the condition
$\mu = 0$ can be consistently imposed in the interacting
theory provided that we restrict ourself to the "light-cone-like"
model manifolds. There are some interesting consequences of this choice.
First of all due to the corner conformal anomaly one gets the
operator insertion at the interaction point of the form
$\exp \gamma\phi(z_i)$. Note that in contrast to the noncritical Polyakov
dual model the operator $:\exp \gamma\phi(z_i) :$ has the conformal
weight different from 1. In fact the coefficient $\gamma$ is a finite
constant uniquely determined by the corner conformal anomaly.
Let us also stress that within the present approach
there is no reason to interpret the integral of
this operator over the world-sheet as the "volume of the universe".

Assuming the same general form of
the transition from the off-shell to the on-shell
amplitudes as in the critical string theory one can expect
that the on-shell amplitudes can be expressed in terms of
correlators of the 2-dim conformal field theory involving
the vertex operators corresponding to the DDF states and the insertions.
This is a very promising feature of the model
-- one can use the techniques of the conformal field theory to
analyse the FCT string amplitudes.

The complete analysis of the interacting FCT string theory
requires solutions of a number of technical and conceptual problems,
which are far beyond the scope of the present paper. We hope however
that the covariant functional integral formulation of the free FCT
string makes the program of constructing the noncritical relativistic
string theory more promising than twenty years ago
and still worth pursuing.
\vspace{1cm}

\section*{Acknowledgements}
We would
like to thank Professor Abdus Salam, the International
Atomic Energy Agency and
UNESCO for  support during our stay at the International
Centre of Theoretical
Physics where the large part of this work was carried out.

\section*{Appendix A}
In this Appendix we gather the results for the corner
conformal anomaly in the case of  Laplace-Beltrami
operator acting on scalar functions with Dirichlet and Neumann
boundary conditions.

The  corner anomaly appears in the expansion ($f$ is a scalar
function)
$${\rm Tr(e}^{t\Delta}f)=\frac{k_{-1}}{t}+\frac{k_{-1/2}}{t^{1/2}}+
k_0+O(t^{1/2})$$
in the t independent part (functional $k_0$) and sums the values of the
function  $f$ in the corners with the appropriate coefficient.
As it was shown by Kac \cite{kac} the contribution from each corner is
independent of the global geometry of the surface. It can be
estimated by mapping a neighborhood of the corner to the wedge on the
plain
with the same opening angle. In the case of Dirichlet boundary conditions
on both arms of the corner with opening angle $\gamma$
Kac derived the following formula for the corner conformal
anomaly
$$
A_{DD}(\gamma)=\frac{\pi^2-\gamma^2}{24\pi\gamma}f(0) \;\;\;.
$$
Using this result one can easily infer the corner conformal anomaly
for the Dirichlet-Neumann and the Neumann-Neumann boundary conditions.
Doubling the corner one gets the following relations
 \begin{eqnarray*}
A_{DD}(\gamma)+A_{DN}(\gamma)&=& A_{DD}(2\gamma)\;\;\;,\\
A_{DN}(\gamma)+A_{NN}(\gamma)&=& A_{NN}(2\gamma)\;\;\;.
\end{eqnarray*}
Using the result of Kac one has
\begin{eqnarray*}
A_{DN}(\gamma)&=& -\ \frac{\pi^2+2\gamma^2}{48\pi\gamma}f(0)\;\;\;,\\
A_{NN}(\gamma)&=& \frac{\pi^2-\gamma^2}{24\pi\gamma}f(0)\;\;\;.
\end{eqnarray*}

In the problems discussed in this paper we have only right angles so we
quote the results for $\gamma=\pi/2$:
\begin{eqnarray*}
A_{DD}(\pi/2)&=& A_{NN}(\pi/2)\;=\;\frac{f(0)}{16}\;\;\;,\\
A_{DN}(\pi/2)&=& -\ \frac{f(0)}{16}\;\;\;.
\end{eqnarray*}

As far as the rectangle with the standard flat metric is concerned
the results above are enough to derive the corner conformal anomaly
for the operators $P^+P, PP^+$ acting on the vector fields
and symmetric traceless tensors, respectively. In this
case the corresponding operators act separately
on every component of vector or tensor fields. Since the components
satisfy independent boundary conditions
the problem can be reduced to the scalar one. The corner conformal
anomaly for these operators has been first derived in \cite{vw}.

\section*{Appendix B}
In this appendix we shall calculate the 1-dim conformal anomaly.
Let $e$ be an einbein on the interval $[0,1]$ and
$$
\Delta_e \equiv -{1\over e}{d\over dt}{1\over e}{d\over dt}\;\;\;,
$$
the 1-dim Laplace operator acting on the space ${\cal S}_N$ of scalar
functions $\psi(t)$ on $[0,1]$ satisfying Neumann boundary conditions
at the ends of $[0,1]$.
Let us denote by ${\cal D}^e\psi$ the functional measure related to
the scalar product on ${\cal S}_N$:
$$
\langle \psi | \psi'\rangle = \int\limits_0^1 e\, dt\,
\overline{\psi(t)}\psi'(t)
\;\;\;.
$$

 The 1-dim conformal anomaly $J[\varphi,\widehat{e}]$ is defined
by the relation
\begin{equation}
{\cal D}^{{\rm e}^{{\varphi \over 2}}\widehat{e}}\psi =
J[\varphi,\widehat{e}]\,
{\cal D}^{\widehat{e}}\psi\;\;\;
\label{conan}
\end{equation}
between the functional
measures corresponding to the einbeins $e={\rm e}^{{\varphi \over 2}}
\widehat{e}$ and $e=\widehat{e}$.
Using the method proposed in the context of the Liouville measure in
2 dimensions \cite{mm} one can derive the following regularized formula
for variation
$$
\delta \log J_N[\varphi,\widehat{e}] = {1\over 4} \lim_{\epsilon \to 0}
\int\limits_0^1 e\,dt\, {\rm e}^{-\epsilon\Delta_{e}}(t,t)\delta \psi(t)
\;\;\;,
$$
where $e = {\rm e}^{{\varphi\over 2}}\widehat{e}$.
In terms of normalized eigenfunctions
$\left\{\psi_m\right\}_{m\geq 0}$ of the operator $\Delta_e$
the formula above takes the form
$$
\delta \log J_N[\varphi,\widehat{e}] ={1\over 4} \lim_{\epsilon \to 0}
\int\limits_0^1 dt \left(
\sum\limits_{m\geq 0} {\rm e}^{-\epsilon {m^2\pi^2\over \alpha^2}}
{\psi_m(t)}^2 \delta\varphi(t) \right)\;\;\;,
$$
where $\alpha = \int_0^1 {\rm e}^{{\varphi \over 2}} \widehat{e} dt$.

Proceeding to the parameterization $s=s(t)$ of $[0.1]$ in which
${\rm e}^{{\varphi \over 2}} \widehat{e} = {\rm const} = \alpha$
and using the expansion
$$
\delta \varphi(s) = \sum\limits_{n\geq 0}\delta\varphi_n \cos\pi ns\;\;\;,
$$
one gets
\begin{eqnarray*}
\delta \log J_N[\varphi,\widehat{e}]
& =& {1\over 4} \lim_{\epsilon \to 0}
\int\limits_0^1 ds
\left(
\sum\limits_{\parbox{26pt}{\scriptsize  $\makebox[1pt]{}n\geq 0$ \\
 $m\geq 1$}}
{\rm e}^{-\epsilon {m^2\pi^2\over \alpha^2}} \delta\varphi_n
 2\cos\pi ns \cos^2\pi ms  + \sum\limits_{n\geq 0}
 \delta\varphi_n \cos\pi ns \right)
 \;\;\;                                  \\
 &=& {1\over 16} \left(
 \delta\varphi(0) + \delta\varphi(1) \right) + \lim_{\epsilon \to 0}
 {1\over 8\sqrt{\pi\epsilon}} \int\limits_0^1
 {\rm e}^{{\varphi \over 2}} \widehat{e} dt
 \,\delta\varphi(t)\;\;\;.
 \end{eqnarray*}
 Integrating with respect to $\varphi$
 one has
 \begin{equation}
\log  J_N[\varphi,\widehat{e}] = {1\over 16} \left(
\varphi(0) + \varphi(1)\right) +
 {1\over 4\sqrt{\pi\epsilon}} \int\limits_0^1
 {\rm e}^{{\varphi \over 2}} \widehat{e} dt\;\;\;.
 \label{jacob}
 \end{equation}
 The corresponding result in the case of Dirichlet
 boundary conditions takes the form
 $$
\log J_D[\varphi,\widehat{e}] = - {1\over 16} \left(
\varphi(0) + \varphi(1)\right) +
 {1\over 4\sqrt{\pi\epsilon}} \int\limits_0^1
 {\rm e}^{{\varphi \over 2}} \widehat{e} dt\;\;\;.
$$
Inserting (\ref{jacob}) into (\ref{conan}) one gets the formula
used in the main text
\begin{equation}
{\cal D}^{{\rm e}^{{\varphi \over 2}}\widehat{e}}\psi =
\exp \left[+{1\over 16} \left(
\varphi(0) + \varphi(1) \right) +
 {1\over 4\sqrt{\pi\epsilon}} \int\limits_0^1
 {\rm e}^{{\varphi \over 2}} \widehat{e} dt \right]
{\cal D}^{\widehat{e}}\psi\;\;\;.
\label{conano}
\end{equation}

\section*{Appendix C}

In this appendix we shall prove the formula (\ref{formula}) of
Subsect.3.2. Consider the mode expansion of the conformal factor
$$
\varphi = {2\over \sqrt{t} } \sum\limits_{m,n \geq 0}
          \varphi_{nm} \cos{\pi n z^0 \over t} \cos \pi m z^1
$$
and the change of variables
\begin{eqnarray}
\psi_{00} &=& \varphi_{00}\;\;\;;\nonumber\\
\psi^+_{k0} &=& \sum\limits_{l\geq k}\varphi_{(2l)0}\;\;\;\;\;,\;\;\;
\psi^-_{k0} \;=\; \sum\limits_{l\geq k}\varphi_{(2l-1)0}
\;\;,\;\;k\geq 1\;\;\;;\label{varia}\\
\psi^-_{km} &=& \sum\limits_{l\geq k}\varphi_{(2l+1)m}\;\;\;,\;\;\;
\psi^-_{km} \;=\; \sum\limits_{l\geq k}\varphi_{(2l+1)m}
\;\;,\;\;k\geq 0,m\geq 1\;\;\;.            \nonumber
\end{eqnarray}
The modes $\widetilde{\varphi}_{im},\widetilde{\varphi}_{fm}$ of
the boundary values $\widetilde{\varphi}_i,\widetilde{\varphi}_f$
of $\varphi$ (\ref{modes}) can be expressed in terms of the variables
(\ref{varia}) as follows
\begin{eqnarray*}
\widetilde{\varphi}_{i0} &=& {2\over \sqrt{t}}
\left( \psi_{00} + \psi^+_{10} + \psi^-_{10} \right)\;\;\;,\\
\widetilde{\varphi}_{f0} &=& {2\over \sqrt{t}}
\left( \psi_{00} + \psi^+_{10} - \psi^-_{10} \right)\;\;\;,\\
\widetilde{\varphi}_{im} &=& {2\over \sqrt{t}}
\left(  \psi^+_{0m} + \psi^-_{0m} \right)\;\;\;,\;\;\;m\geq 1\;\;\;,\\
\widetilde{\varphi}_{fm} &=& {2\over \sqrt{t}}
\left(  \psi^+_{0m} - \psi^-_{0m} \right)\;\;\;,\;\;\;m\geq 1\;\;\;.
\end{eqnarray*}
In terms of the variables (\ref{varia}) the l.h.s. of the formula
(\ref{formula}) can be written as the iterated integral
\begin{eqnarray}
Z&=& {\rm const} \int d\psi_{00} \int \prod\limits_{k\geq 1}
d\psi^+_{k0}d\psi^-_{k0} \nonumber\\
&\times& \exp \left[- {b\over 2t^2}
\sum\limits_{k\geq1} 4k^2 \left(\psi^+_{k0} -\psi^+_{(k+1)0}\right)^2
\right]
\label{inte} \\
&\times& \exp \left[- {b\over 2t^2}
\sum\limits_{k\geq1} (2k+1)^2 \left(\psi^-_{k0} -\psi^-_{(k+1)0}\right)^2
\right] \times Z_1\;\;\;,\nonumber
\end{eqnarray}
where
\begin{eqnarray}
Z_1
&=&
\int \prod\limits_{\parbox{26pt}{\scriptsize
\makebox[1pt]{} $k\geq 0$ \\ $m\geq 1$}}
d\psi^+_{km}d\psi^-_{km}
 \exp \left[ -b \sum\limits_{m\geq1}
m^2 \left( \psi^+_{0m} -\psi^+_{1m} \right)^2 \right] \nonumber\\
&\times& \exp \left[    -{b\over 2}
 \sum\limits_{\parbox{26pt}{\scriptsize
\makebox[1pt]{} $k\geq 1$ \\ $m\geq 1$}}
 \left( {4k^2\over t^2} +m^2 \right)
\left( \psi^+_{km} - \psi^+_{(k+1)m} \right)^2 \right]
\label{inte2}\\
&\times& \exp \left[ - {b\over 2}
\sum\limits_{\parbox{26pt}{\scriptsize
\makebox[1pt]{} $k\geq 0$ \\ $m\geq 1$}}
\left( \frac{(2k+1)^2}{t^2} + m^2 \right)
\left( \psi^-_{km} - \psi^-_{(k+1)m} \right)^2 \right] \\
&\times&
F[\widetilde{\varphi}_i,\widetilde{\varphi}_f] \;\;\;,\nonumber
\end{eqnarray}
end
$$
b= {25-d\over 48}\pi\;\;\;.
$$
Let us note that the functional
$F[\widetilde{\varphi}_i,\widetilde{\varphi}_f]$
is independent of the modes $\left\{ \psi^+_{km},\psi^-_{km}
\right\}_{k \neq 0}$ and
$Z_1\equiv Z_1(\widetilde{\varphi}_{i0},\widetilde{\varphi}_{f0})$
is a function only of the zero modes of
$\widetilde{\varphi}_i,\widetilde{\varphi}_f$.

The integrals $Z,Z_1$ involve chains of Gaussian integrals which can
be exactly calculated by means of the formula ($\lambda_{N+1} =0$)
\begin{eqnarray}
\int \prod\limits_{k\geq 1}^N
\exp  \!\!\!\!\!\!\!\!& &\!\!\!\!\!\left[ -
{\beta \over 2} \lambda_0 (x_0 -x_1)^2
-{\beta \over 2} \sum\limits_{k\geq 1}^N \lambda_k
(x_k - x_{k+1})^2 \right] =\nonumber\\
&=& \left( \prod\limits_{k\geq 0}^N \frac{\beta \lambda_k}{2\pi}
    \right)^{-{1\over 2}} \left( {\beta\over 2\pi A} \right)^{{1\over 2}}
    \exp \left(-{\beta \over 2\pi A} x_0^2 \right)
    \;\;\;,\label{gaus}
\end{eqnarray}
where
$$
A = \sum\limits_{k\geq 0}^N {1 \over \lambda_k}\;\;\;\;\;.
$$

Applying the formula (\ref{gaus}) to the integral $Z$
one obtains
\begin{eqnarray*}
Z\;=\;&{\rm const}& \left[ \prod\limits_{k\geq 1}
{4k^2\over t^2}\frac{(2k+1)^2}{t^2} \right]^{-{1\over 2}}
\int d\psi_{00} \int \frac{d\psi^+_{10}}{t\sqrt{A^+_0}}
\int \frac{d\psi^-_{10}}{t\sqrt{A^-_0}} \\
&\times& \exp \left[ -\frac{b}{t^2A^+_0}{\psi^+_{10}}^2
-\frac{b}{t^2A^-_0}{\psi^-_{10}}^2
\right] \\ &\times&
Z_1\left({2\over \sqrt{t}}
\left( \psi_{00} + \psi^+_{10} + \psi^-_{10} \right),
{2\over \sqrt{t}}
\left( \psi_{00} + \psi^+_{10} - \psi^-_{10} \right)\right)
\;\;\;,
\end{eqnarray*}
where
\begin{eqnarray*}
A^+_0 &=& \sum\limits_{k\geq 1} {1\over 4k^2} \;=\;{\pi^2\over 24}
\;\;\;,\\
A^-_0 &=& \sum\limits_{k\geq 0} {1\over (2k+1)^2} \;=\;{\pi^2\over 8}
\;\;\;.
\end{eqnarray*}
Changing variables
\begin{eqnarray*}
\widetilde{\varphi}_{i0} &=& {2\over \sqrt{t}} \left(
\psi_{00} +\psi^+_{10} +\psi^-_{10} \right)\;\;\;,\\
\widetilde{\varphi}_{f0} &=& {2\over \sqrt{t}} \left(
\psi_{00} +\psi^+_{10} -\psi^-_{10} \right)\;\;\;,\\
\varphi_{00} &=& \psi_{00}\;\;\;,
\end{eqnarray*}
and integrating over $\varphi_{00}$ one finally gets
\begin{equation}
Z= {\rm const}\; t^{-{1\over 2}} \int
d\widetilde{\varphi}_{i0} d\widetilde{\varphi}_{f0}
\exp \left[ - {b\over 2\pi^2}\frac{\left(
\widetilde{\varphi}_{f0} - \widetilde{\varphi}_{i0} \right)^2}{t}
\right]
Z_1(\widetilde{\varphi}_{i0},\widetilde{\varphi}_{f0})\;\;.
\label{zet}
\end{equation}
In the case of the integral $Z_1$ the formula (\ref{gaus})
yields
\begin{eqnarray*}
Z_1 \;=\;&{\rm const}& \left[
\prod\limits_{\parbox{26pt}{\scriptsize
\makebox[1pt]{} $k\geq 0$ \\ $m\geq 1$}}
\left( {4k^2\over t^2} + m^2\right)
\left(\frac{(2k+1)^2}{t^2} +m^2\right) \right]^{-{1\over 2}}
\int \prod\limits_{m\geq 1} \frac{d\psi^+_{0m}}{t\sqrt{A^+_m}}
\frac{d\psi^-_{0m}}{t\sqrt{A^-_m}} \\
&\times& \exp \left[ -{b\over 2}
\sum\limits_{m\geq 1} \left( \frac{{\psi^+_{0m}}^2}{A^+_m}
+ \frac{{\psi^-_{0m}}^2}{A^-_m} \right)
\right]
\times
F[\widetilde{\varphi}_i,\widetilde{\varphi}_f] \;\;\;,
\end{eqnarray*}
where
\begin{eqnarray*}
A^+_m &=& {1\over 2m^2} + \sum\limits_{k\geq 1}
\frac{t^2}{4k^2 + t^2m^2}\;=\;{\pi t\over 4m}
\coth {\pi mt\over 2}\;\;\;,\\
A^-_m &=&  \sum\limits_{k\geq 1}
\frac{t^2}{(2k+1)^2 + t^2m^2}\;=\;{\pi t\over 4m}
\tanh {\pi mt\over 2}\;\;\;.
\end{eqnarray*}
Changing variables
\begin{eqnarray*}
\widetilde{\varphi}_{im} &=& {2\over \sqrt{t}}
\left(\psi^+_{0m} +\psi^-_{0m} \right)\;\;\;,\\
\widetilde{\varphi}_{fm} &=& {2\over \sqrt{t}}
\left(\psi^+_{0m} -\psi^-_{0m} \right)\;\;\;   ,
\end{eqnarray*}
and using the formula
$$
\eta(t) =     \prod\limits_{\parbox{26pt}{\scriptsize
\makebox[1pt]{} $k\geq 0$ \\ $m\geq 1$}}
\left({k^2\over t^2} +m^2 \right)\;\;\;,
$$
one has
\begin{eqnarray}
Z_1 \;=\;&{\rm const}& \eta(t)^{-{1\over 2}} \int
\prod\limits_{m\geq 1} d\widetilde{\varphi}_{im}\;
d\widetilde{\varphi}_{fm} \nonumber\\
&\times& \exp \left[ -{b\over 4\pi^2}
\sum\limits_{m\geq 1}\frac{\pi m}{\sinh \pi mt}
\left[ \left( \widetilde{\varphi}_{im}^2 +
\widetilde{\varphi}_{fm}^2 \right) \cosh \pi mt
- 2\widetilde{\varphi}_{im}\widetilde{\varphi}_{fm}
\right]\right]\nonumber\\
&\times&
F[\widetilde{\varphi}_i,\widetilde{\varphi}_f]\;\;\;.
\label{zett}
\end{eqnarray}
Substituting (\ref{zett}) to (\ref{zet}) one gets the generalized
Forman formula (\ref{formula}).

\end{document}